\documentclass[pra,twocolumn,superscriptaddress,floatfix]{revtex4}
\usepackage{graphicx,amsfonts,amssymb,amsmath,hyperref,enumerate}

\usepackage{tikz}
\usetikzlibrary{shapes.geometric,patterns,positioning,matrix}
\usetikzlibrary{shadows,calc,3d,arrows.meta,decorations.markings,decorations.pathreplacing}

\usepackage{pgffor}
\usepackage{ifthen}
\usepackage{tikz-3dplot}

\usepackage{siunitx}
\usepackage{float}

\definecolor{orange}{rgb}{1,0.5,0}
\definecolor{purple}{rgb}{0.5,0,0.5}

\definecolor{TCBlue}{HTML}{56ADFF}
\definecolor{TCAntiBlue}{HTML}{ADFF56}
\definecolor{lightblue}{HTML}{1E90FF}
\definecolor{lightgreen}{HTML}{00FF00}

\definecolor{TensorBlue}{HTML}{56ADFF}
\definecolor{TensorBlueOne}{HTML}{2394FF}
\definecolor{TensorBlueTwo}{HTML}{89C6FF}

\definecolor{TensorGreen}{HTML}{9AEE68}
\definecolor{TensorGreenOne}{HTML}{7BE93A}
\definecolor{TensorGreenTwo}{HTML}{B9F396}
\definecolor{TensorAwesomeGreen}{HTML}{53C061}

\definecolor{TensorPink}{HTML}{EE689A}
\definecolor{TensorPinkOne}{HTML}{E93A7B}
\definecolor{TensorPinkTwo}{HTML}{F396B9}

\definecolor{gray1}{HTML}{cccccc}
\definecolor{gray2}{HTML}{999999}
\definecolor{gray3}{HTML}{666666}
\definecolor{gray4}{HTML}{323232}

\definecolor{UniRed}{HTML}{BD0927}

\newif\ifhyper
\hypertrue
\ifhyper
\hypersetup{
   citecolor = {green},
   colorlinks = {true}, 
   urlcolor = {blue} 
}
\fi

\newcommand{\beq}{\begin{equation}}
\newcommand{\eeq}{\end{equation}}
\newcommand{\beqa}{\begin{eqnarray}}
\newcommand{\eeqa}{\end{eqnarray}}

\def\ri{\mathrm i}

\def\re{\mathrm e}

\def\Longarrow{\protect\@lra}
\def\@lra{\relbar\joinrel\relbar\joinrel\relbar\joinrel%
          \relbar\joinrel\rightarrow}

\usepackage{scalerel}

\def\vecsign{\mathchar"017E}
\def\dvecsign{\smash{\stackon[-1.95pt]{\vecsign}{\rotatebox{180}{$\vecsign$}}}}
\def\dvec#1{\def\useanchorwidth{T}\stackon[-4.2pt]{#1}{\,\dvecsign}}
\usepackage{stackengine}
\stackMath

\usepackage{graphicx,accents}

\makeatletter
\DeclareRobustCommand{\cev}[1]{%
  \mathpalette\do@cev{#1}%
}
\newcommand{\do@cev}[2]{%
  \fix@cev{#1}{+}%
  \reflectbox{$\m@th#1\vec{\reflectbox{$\fix@cev{#1}{-}\m@th#1#2\fix@cev{#1}{+}$}}$}%
  \fix@cev{#1}{-}%
}
\newcommand{\fix@cev}[2]{%
  \ifx#1\displaystyle
    \mkern#23mu
  \else
    \ifx#1\textstyle
      \mkern#23mu
    \else
      \ifx#1\scriptstyle
        \mkern#22mu
      \else
        \mkern#22mu
      \fi
    \fi
  \fi
}

\makeatother

\begin{document}

\title{Quantum criticality on a chiral ladder: an $SU(2)$ iDMRG study}

\author{Philipp Schmoll}
\affiliation{Institute of Physics, Johannes Gutenberg University, 55099 Mainz, Germany}
\affiliation{Graduate School Materials Science in Mainz, Staudingerweg 9, 55128 Mainz, Germany}

\author{Andreas Haller}
\affiliation{Institute of Physics, Johannes Gutenberg University, 55099 Mainz, Germany}
\affiliation{Graduate School Materials Science in Mainz, Staudingerweg 9, 55128 Mainz, Germany}

\author{Matteo Rizzi}
\affiliation{Institute of Physics, Johannes Gutenberg University, 55099 Mainz, Germany}
\affiliation{Institute of Quantum Control (PGI-8), Forschungszentrum J\"ulich, D-52425 J\"ulich, Germany}
\affiliation{Institute for Theoretical Physics, University of Cologne, D-50937 K\"oln, Germany}

\author{Rom\'an Or\'us}
\affiliation{Institute of Physics, Johannes Gutenberg University, 55099 Mainz, Germany}
\affiliation{Donostia International Physics Center, Paseo Manuel de Lardizabal 4, E-20018 San Sebasti\'an, Spain}
\affiliation{Ikerbasque Foundation for Science, Maria Diaz de Haro 3, E-48013 Bilbao, Spain}

\begin{abstract}

In this paper we study the ground state properties of a ladder Hamiltonian with chiral $SU(2)$-invariant spin interactions, a possible first step towards the construction of truly two dimensional non-trivial systems with chiral properties starting from quasi-one dimensional ones. Our analysis uses a recent implementation by us of $SU(2)$ symmetry in tensor network algorithms, specifically for infinite Density Matrix Renormalization Group (iDMRG). After a preliminary analysis with Kadanoff coarse-graining and exact diagonalization for a small-size system, we discuss its bosonization and recap the continuum limit of the model to show that it corresponds to a conformal field theory, in agreement with our numerical findings. In particular, the scaling of the entanglement entropy as well as finite-entanglement scaling data show that the ground state properties match those of the universality class of a $c = 1$ conformal field theory (CFT) in $(1+1)$ dimensions. We also study the algebraic decay of spin-spin and dimer-dimer correlation functions, as well as the algebraic convergence of the ground state energy with the bond dimension, and the entanglement spectrum of half an infinite chain. Our results for the entanglement spectrum are remarkably similar to those of the spin-$1/2$ Heisenberg chain, which we take as a strong indication that both systems are described by the same CFT at low energies, i.e., an $SU(2)_1$ Wess-Zumino-Witten theory. Moreover, we explain in detail how to construct Matrix Product Operators for $SU(2)$-invariant three-spin interactions, something that had not been addressed with sufficient depth in the literature.

\end{abstract}

\maketitle

\section{Introduction}
\label{sec1}
The study of quantum criticality with Density Matrix Renormalization Group \cite{DMRG} has a long history. As such, quantum critical systems have an infinite correlation length, and it is well-known that this cannot be exactly captured by DMRG, which is based on Matrix Product States \cite{MPS}, but perhaps rather by other tensor networks \cite{TN} such as the Multiscale Entanglement Renormalization Ansatz \cite{MERA} and Tree Tensor Networks \cite{TTN}. Nevertheless, DMRG is very efficient and simple to program, and this is the reason why often it is the preferred option to study criticality, both in its finite-size and infinite-size (iDMRG) \cite{iDMRG} versions. The approach, then, is to push forward as much as possible the MPS bond dimension, and do appropriate finite-size and/or finite-entanglement \cite{finent} scalings to extract critical properties. To push the bond dimension, one of the best ideas is to implement symmetries. In particular, for $SU(2)$-invariant systems, the use of $SU(2)$ at the level of the MPS has proven remarkably useful in simulations of, e.g., Heisenberg quantum spin chains and quasi-1d systems \cite{su2dmrg}.

In this paper we use our own implementation of $SU(2)$-invariant iDMRG \cite{ourSU2} to study the ground-state properties of a spin-1/2 2-leg ladder with chiral 3-spin interactions. The model is similar to the system in Ref.~\cite{2dchiral}, defined on the two-dimensional Kagome lattice and with a Hamiltonian made of purely-chiral 3-spin terms \footnote{Here we use the word ``chiral" in the sense that the Hamiltonian is not even under a time-reversal operation. As we will see later, the continuum limit is a field theory which is odd under time-reversal.}. In that model, a ground state analysis using 2d DMRG in cylinders unveiled a ground state with chiral topological order.  Here, our motivation for studying the ladder is multi-fold. First, it allows us to study the crossover from 1d to 2d for chiral interactions. In particular, we find that the ladder has chiral properties similar to those of the chiral edge mode in the 2d model, and a critical ground state with a central charge $c=1$ and other critical exponents that we characterize numerically. Our findings are also compatible with previous studies showing that the continuum limit of the model is a $(1+1)$-dimensional conformal field theory \cite{continuum1, continuum2, continuum3}, and show in particular that the entanglement spectrum matches the expected behaviour of an $SU(2)_1$ Wess-Zumino-Witten (WZW) theory at low energies. Moreover, the technical simulation of the model allows us to understand how to implement $SU(2)$-invariant Matrix Product Operators (MPO) for Hamiltonians with 3-spin chiral interactions: something that, to our surprise, had not yet been discussed with enough detail in the literature. Finally, the model is one of the simplest $SU(2)$-generalizations of a quite widespread strategy in trying to access non-trivial two dimensional systems with chiral properties, starting from quasi-one dimensional ones (often dubbed as ``wire deconstructionism'')~\cite{wireconstructionism,otherrefs}: the key idea being to gap out right movers of a wire with left movers (or vice versa) of the neighbouring one by means of suitable interactions, remaining at the end with a chiral edge current on the external legs of the ladder. Such an approach is experiencing a growing application in cold atoms, photonics and nanowire experiments~\cite{coldatoms}. By exploring the properties of our ladder model we provide further intuition about the structure of such 2d chiral phases.

The structure of this paper is as follows. In Sec.~\ref{sec2} we first introduce the details of the model Hamiltonian for the chiral ladder with 3-spin interactions. Then we explain briefly the expected behaviour from Kadanoff coarse-graining and small-size exact diagonalization results, before we discuss its bosonization. In Sec.~\ref{sec3} we explain some details about the implementation of our numerical method, namely, $SU(2)$-invariant iDMRG. In Sec.~\ref{sec4} we present the results of our simulation, where we show that the ground state of the system corresponds to a conformal field theory (CFT) with $c=1$. We additionally study the algebraic decay of spin-spin and dimer-dimer correlation functions, as well as the entanglement spectrum and the convergence of the ground state energy. Our results for the entanglement spectrum are remarkably similar to those of the spin-$1/2$ Heisenberg chain, which we take as a strong indication that the CFT at low energies is a $SU(2)_1$ WZW theory. We wrap up our conclusions in Sec.~\ref{sec5}. Finally, in Appendix we compute the spin-current operators (Appendix~\ref{append1}), review the continuum limit of the model by Huang et al. presented in Ref.~\cite{continuum1} (Appendix~\ref{append2}) and explain the details of how to construct the $SU(2)$-invariant MPO for the Hamiltonian that we want to simulate (Appendix~\ref{append3}), focusing on chiral 3-spin interactions. In Appendix~\ref{append4} we provide numerical data on the finite-entanglement scaling of the entanglement spectrum.

\section{Chiral ladder}
\label{sec2}

\subsection{The model}

\begin{figure}
	\includegraphics[width=\columnwidth]{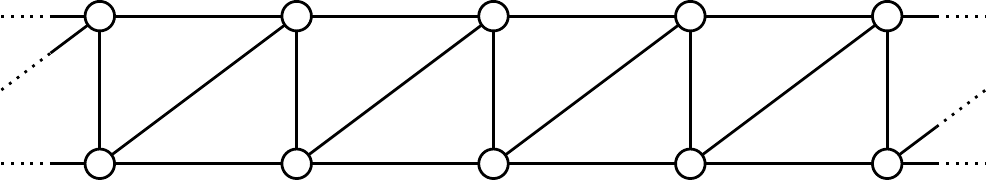}
	\caption{2-leg ladder made of triangles, for the model in Eq.~(\ref{chiralH}).}
	\label{fig1}
\end{figure}

\begin{figure}
	\centering
	\includegraphics[width=\columnwidth]{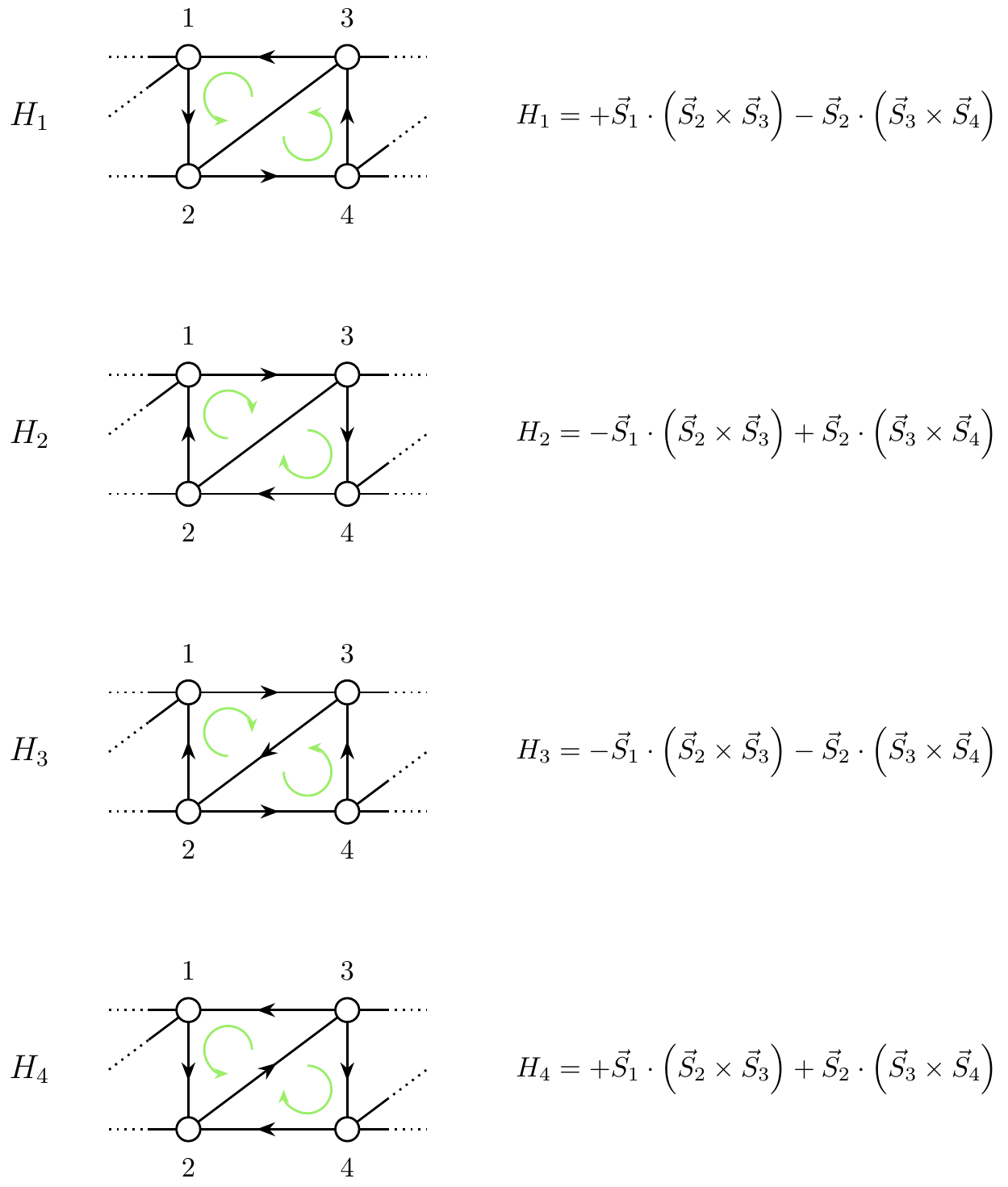}
	\caption{Different orientations of the chiral triple product result in different models. In the first two cases the orientation is chosen to be the same whereas it is opposite in the last two cases.}
	\label{fig2}
\end{figure}

The model that we analyze in this paper is a 2-leg ladder with chiral interactions on triangles. Specifically, it is a model of spins-$1/2$ on the sites of the ladder of Fig.~\ref{fig1} via the three-spin interaction Hamiltonian
\beq
    H = \sum_i J_{i} ~\mathbf{S}_i \cdot \left(  \mathbf{S}_{i+1} \times \mathbf{S}_{i+2} \right),
    \label{chiralH}
\eeq
with $ \mathbf{S}_i$ the spin-$1/2$ operator at site $i$. The sites of the ladder are labeled in a snake-like pattern as shown in the details of the ladder in Fig.~\ref{fig2} and both triangles follow this snake-like labelling (upper triangles 1-2-3, lower triangles 2-3-4). We will consider the cases where $J_{i} \in \{\pm 1\}$ in which the coupling coefficients depend on the traversal of the triangle. A triangle formed by sites $i,i+1,i+2$ is traversed in (against) the direction of the labels if $J_{i} = 1$ ($J_{i} = -1$). This can be rephrased to clockwise or anticlockwise configurations for each triangle. The triangles of the full ladder are all clockwise configured if $J_{i} = (-1)^i$ (anticlockwise if $J_{i} = -(-1)^i$). Mixing the two scenarios gives rise to $J_i =1\ \forall\, i$ ($J_i = -1\ \forall\, i$), which leads to a staggered, anticlockwise/clockwise (clockwise/anticlockwise) configuration pattern.

Playing with different clockwise/anticlockwise configurations of the triangles, we can get different Hamiltonians. For instance, for a unit cell of two triangles we can get the four configurations presented in Fig.~\ref{fig2}. In the figure, two of the configurations ($H_1$ and $H_2$) have the same orientation of the triangles (i.e., both clockwise or both anticlockwise), and two  ($H_3$ and $H_4$) have opposite orientation (i.e., one clockwise and one anticlockwise). Since we have $H_1 = -H_2$ and $H_3 = -H_4$, both pairs of Hamiltonians have the same energy spectrum. Therefore, for the physical properties only the relative orientation between the two triangles matters and it is sufficient to restrict to $H_1$ and $H_3$ as different cases.

Both Hamiltonians are odd under time-reversal symmetry (${\bf S}_i\rightarrow -{\bf S}_i$), which results in ${\mathcal T H_i \mathcal T^{-1} = - H_i}$. The combination of two mirror symmetries (which is equivalent to an inversion at the chain center) leaves Hamiltonian $H_1$ invariant, whereas $H_3$ transforms as ${\mathcal P H_3 \mathcal P^{-1} = -H_3}$. This  main difference between the two cases with different relative triangle orientations ($H_1$ and $H_3$) results in different behaviour of the edge states: while edge states for $H_1$ are expected to be counter-propagating, they propagate in the same direction for $H_3$, see the arrows in Fig.~\ref{fig2}. In what follows we show that this intuition is indeed true.

\subsection{First intuition with Kadanoff coarse-graining}

The first approach we take to understand the dominant physics of the model consists in a Kadanoff-like coarse-graining procedure of the triangles into effective spin-1/2's. In particular, we simply (i) project the $2^3$-dimensional Hilbert space of the triangle $n$ starting at site $i=3n-2$ onto the $2$-dimensional subspace of lowest energy via the isometry $W_n:  \tfrac{1}{2} \otimes \tfrac{1}{2} \otimes \tfrac{1}{2}  \longrightarrow \tfrac{1}{2}$, (ii) construct the representation of the operators $W_{n} \mathbf{S}_{j} W_{n}^\dagger$ and $W_{n} \left(\mathbf{S}_j \times \mathbf{S}_{j+1}\right) W_{n}^\dagger$ in this subspace, and then (iii) look for the emerging Hamiltonian.

The first step is rather easy, once we recall that the $SU(2)$-invariant Hamiltonian triangle term of Eq.~\eqref{chiralH} has to be proportional to the identity in the different subspaces with definite total spin, that it has null trace and that it will be vanishing once the three spins are all parallel arranged. Indeed, a bit of algebra with Pauli matrices and Levi-Civita symbols leads to the expression:
\begin{equation}
\mathbf{S}_1 \cdot \left(\mathbf{S}_2 \times \mathbf{S}_3\right) =
	\sum_{\alpha=\pm}  \alpha \frac{\sqrt{3}}{4} \mathbb{P}_{1/2,\alpha} + 0 \, \mathbb{P}_{3/2} ,
\end{equation}
where $\tfrac{1}{2} \otimes \tfrac{1}{2} \otimes \tfrac{1}{2} = \tfrac{1}{2}_+ \oplus \tfrac{1}{2}_- \oplus \tfrac{3}{2}$ and $\mathbb{P}$ are the corresponding projectors. $\frac12_\pm$ are the subspaces of the spin-1/2 states with positive and negative energy. The searched isometry will then be depending on the sign of the triangle coupling, i.e.,
\begin{equation}
W_n W_n^\dagger = \mathbb{P}_{1/2, - \mathrm{sgn}(J_{3n-2})} \ .
\end{equation}

Next we have to construct the coarse-grained expressions of the spin operators involved in the interaction between triangles $n$ and $n+1$.
It turns out that we can choose the projectors such that $\forall j \in \{3n-2, 3n-1, 3n\}$ and $\alpha\in\{\pm\}$:
\begin{eqnarray}
W_{1/2,\alpha} \mathbf{S}_{j} W_{1/2,\alpha}^\dagger & = & \frac{1}{3} \mathbf{\widetilde{S}}_n
\\
W_{1/2,\alpha} \left( \mathbf{S}_{j} \times \mathbf{S}_{j+1} \right) W_{1/2,\alpha}^\dagger & = &  \frac{\alpha}{\sqrt{3}}\,  \mathbf{\widetilde{S}}_n
\end{eqnarray}
with $\mathbf{\widetilde{S}}_n$ the new effective spin 1/2.
The resulting effective Hamiltonian then reads:
\begin{equation}\label{eq:effham}
H_{\mathrm{eff}} =
- \mathrm{sgn}\left(J_1 J_2\right) \frac{|J_1| + |J_2|}{3\sqrt{3}}  \sum_{n=1}^N \mathbf{\widetilde{S}}_n \cdot \mathbf{\widetilde{S}}_{n+1}
\ ,
\end{equation}
where $N \simeq L/3$ is the total number of effective triangles,
and we neglected an additive term $- \tfrac{\sqrt{3}}{4} \tfrac{|J_1| + |J_2|}{2} N$.
We thus obtained an emerging spin-1/2 Heisenberg chain, whose magnetic character (ferro- or antiferro-) depends on the mutual signs of the triangle couplings $J_1$ and $J_2$, and we can resort to a wealth of known facts to foresee the behaviour of our triangle ladder.

If the triangles are all (anti-)clockwise oriented ($H_1$ and $H_2$), then the effective model~\eqref{eq:effham} is anti-ferromagnetic: we therefore predict that it will be gapless, with central charge $c=1$, and that its ground state would tend to minimize the total-spin of the chain, i.e., for even $N$ will be in the zero total spin sector.
Conversely, if the triangles have mixed character ($H_3$ and $H_4$), then the effective model~\eqref{eq:effham} is ferromagnetic: we have thus good reasons to expect that the system will try to maximize its total spin, giving rise to a macroscopic degeneracy of the ground state manifold, and without a well-defined CFT character.  Of course, such low-energy projection is a very strong simplification and further corrections would be needed to describe the full richness of the model (e.g., the degeneracy counting of the case $J_1 = J_2$ in finite systems will be non-trivial). But still, we will see below that the main results obtained by this simple analysis are in fact  confirmed by more sophisticated theoretical and numerical approaches.

\subsection{Exact diagonalization of small systems}

The intuition obtained from the Kadanoff blocking in the previous section can be further corroborated by a simple exact diagonalization exercise. Specifically, here we perform exact diagonalization for small sizes, in particular for 16 spins. For this case, we compute the ground state and low-energy excited states and evaluate some observables. Of particular interest in order to assess chirality are spin-current operators of the form
\begin{align}
  \begin{split}
      \mathcal J_{i,i+1}^z &= - J_{i-1} S_{i-1}^z \left( \boldsymbol S_{i} \boldsymbol S_{i+1} \right) - J_i \left( \boldsymbol S_{i} \boldsymbol S_{i+1} \right) S_{i+2}^z \\
      \mathcal J_{i,i+2}^z &= - J_i \left( \boldsymbol S_{i} \boldsymbol S_{i+2} \right) S_{i+1}^z
  \end{split}
\end{align}
which describe the flow of the $z$-component of magnetization from sites $i$ to site $i+1$ and $i+2$ respectively (notice that, by SU(2) invariance, there is not a preferred spin-component). $\mathcal J_{i,i+1}^z$ measures the currents on the rung and slash links, to which there are contributions from two triangles. $\mathcal J_{i,i+2}^z$ measures the currents in the chains with only a single triangle contribution. The current operators can be derived by taking the commutator of the spin operator and the Hamiltonian, which is presented in Appendix~\ref{append1} for a general $N$-leg ladder. For the wave functions obtained from our small-size exact diagonalization, we evaluate the expectation value of this current operator for the \emph{up, down, rung} and \emph{slash} pairs of sites. Notice that in the chosen basis described below there are no $\mathcal J^x$ and $\mathcal J^y$ current components, so that the plotted current $\mathcal J^z$ is the total current in the system.

\begin{figure}
  \centering
  \includegraphics[width=0.95\columnwidth]{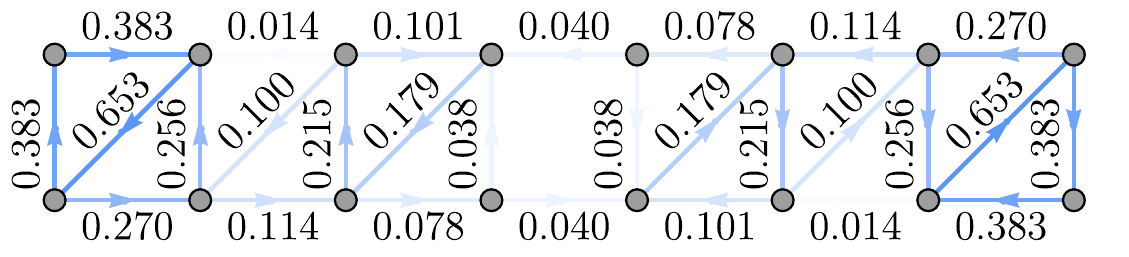}
  \includegraphics[width=0.95\columnwidth]{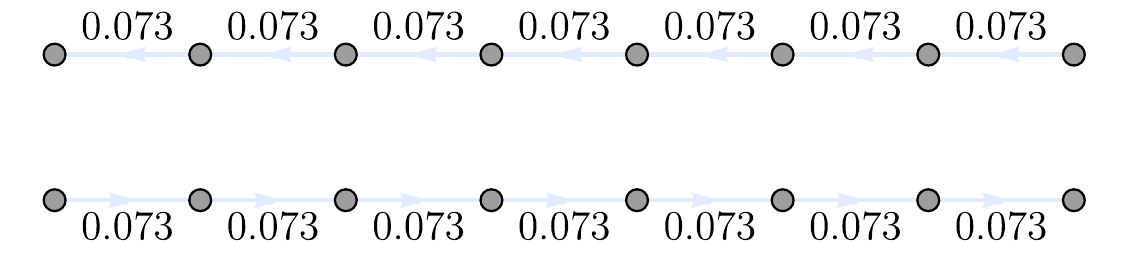}
  \caption{Expectation values of the link currents for the first excited state of $H_1$ with 16 spins and $\langle {\bf S}^2 \rangle = 2$, $S^z = -1$ for open boundary conditions (top) and periodic boundary conditions (bottom). The color refers to the strength of the currents normalized to the maximal current $\mathcal J^{\rm max} = 0.107$ in Figs.~\ref{figpat1} and \ref{figpat2}. Both cases have counter-propagating edge currents, with periodic boundary conditions showing translation invariance and no currents on the rung and slash links. The first excited state with $\langle {\bf S}^2 \rangle = 2$ and $S^z = 0$ does not show $\mathcal J^z$ current expectation values, for $S^z = +1$ the current patterns are inverted.}
  \label{figpat1}
\end{figure}

For the Hamiltonian configuration $H_1$ (or equivalently $H_2$), we find from our results that the ground state is a singlet of $SU(2)$ with total spin zero, i.e., $\langle {\bf S}^2 \rangle = 0$ with ${\bf S}$ the total spin vector operator. Thus, we find that the ground state does not carry any currents. However, in the first excited state (an $SU(2)$ triplet) the pattern of currents for every pair of sites corresponds to the one in Fig.~\ref{figpat1}, where we show both open and periodic boundary conditions. The chirality of the currents in the bulk is clear, and matches the intuition from Fig.~\ref{fig2}.

\begin{figure}
  \centering
  \includegraphics[width=0.95\columnwidth]{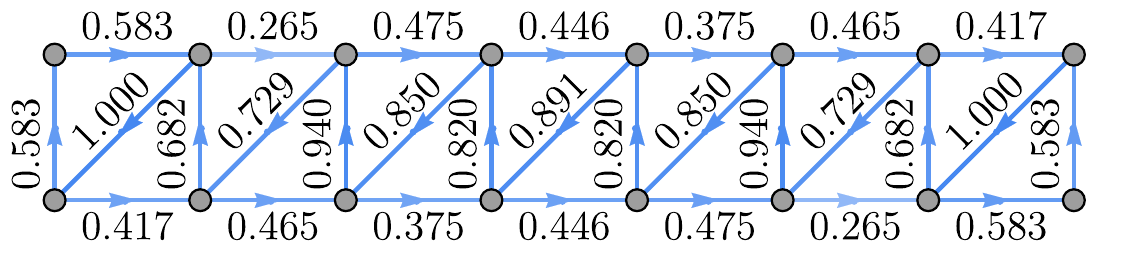}
  \includegraphics[width=0.95\columnwidth]{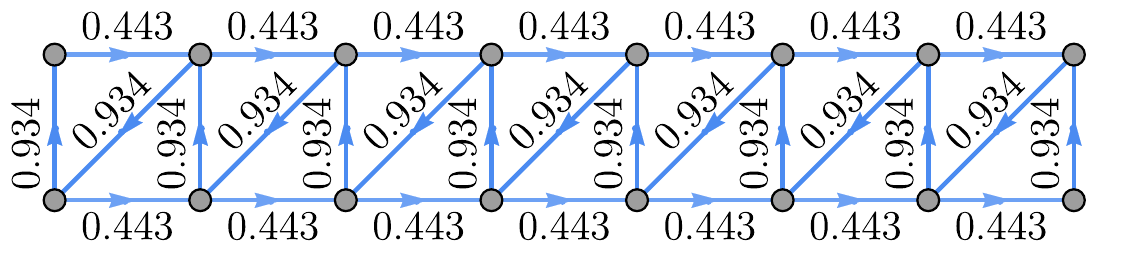}
  \caption{Expectation values of the link currents for the ground state of $H_3$ with 16 spins and $\langle {\bf S}^2 \rangle = 12$, $S^z = -3$ for open boundary conditions (top) and periodic boundary conditions (bottom). The color refers to the strength of the currents normalized to the maximal current $\mathcal J^{\rm max} = 0.107$ in Figs.~\ref{figpat1} and \ref{figpat2}. Both cases have co-propagating edge currents, with periodic boundary conditions showing translation invariance and no oscillations in the strength of the currents. The ground state with $\langle {\bf S}^2 \rangle = 12$ and $S^z = 0$ does not show $\mathcal J^z$ current expectation values, for $S^z = +3$ the current patterns are inverted.}
  \label{figpat2}
\end{figure}

Complementary, for the Hamiltonian configuration $H_3$ (or equivalently $H_4$), we find that the ground state has a well-defined total spin and is also degenerate, according to the data in Table~\ref{tab1}. In this case, we diagonalize the subspace of degenerate ground states in the ${\bf S}^2$ and the $S^z$ basis and pick states with fixed total and $z$-component of the spin. We find that in such ground states there is a non-trivial current behaviour, as shown in Fig.~\ref{figpat2}. Again, the observed pattern also matches our intuitive picture from Fig.~\ref{fig2}.

\begin{table}
	\centering
	\begin{tabular}{||c|c|c||}
	\hline
	~~\# Spins~~ & ~~Degeneracy~~ & ~~Total spin~~  \\
         \hline
         \hline
4 & 3 & 1 \\
6 & 3 & 1 \\
8 & 5 & 2 \\
10 & 5 & 2 \\
12 & 7 & 3 \\
14 & 7 & 3 \\
16 & 7 & 3 \\
18 & 9 & 4 \\
20 & 9 & 4 \\
22 & 11 & 5 \\
24 & 11 & 5 \\
\hline
\end{tabular}
\caption{Ground state degeneracy and total spin for different lengths, for configurations $H_3$ and $H_4$.}
\label{tab1}
\end{table}

From our small-size study with exact diagonalization we learn a couple of important things in order to study this model. First, configurations $H_3$ and $H_4$ do \emph{not} have counter-propagating chiral edge modes and, moreover, have a \emph{covariant} ground state with a well-defined, non-zero, total spin $S$ -- therefore the ground manifold is a $(2S+1)$-plet. Second, configurations $H_1$ and $H_2$  have counter-propagating chiral edge modes as well as a singlet ground state, which is thus $SU(2)$ invariant.

As a matter of fact, the reason for the ground state of $H_3$ breaking $SU(2)$ symmetry down to $U(1)$ lies in the different overall character of the Hamiltonian: as readily visible on short chains, $H_1$ is globally anti-ferromagnetic, while $H_3$ is instead ferromagnetic. This means, practically, that the ground state of $H_1$ belongs to the singlet sector (total spin zero), while the energy for $H_3$ would be minimized by a state with large spin. If we do not make use of symmetries, the multiplicity of the $H_3$ ground state manifold diverges at the targeted thermodynamic limit, thus making it difficult for numerical algorithms to converge.

Given the above, in this paper we choose to analyze in detail the ground state of configuration $H_1$ (or equivalently $H_2$), which is an $SU(2)$ singlet, with an $SU(2)$-invariant infinite-DMRG code. The case of configuration $H_3$ (and $H_4$) could be better assessed by an MPS  code that incorporates $U(1)$ symmetry instead, and/or an $SU(2)$-code that can target generic covariant states. Therefore we focus here entirely on the  configuration providing an $SU(2)$-invariant ground state.

\subsection{Bosonization}

A wide class of interactions can be treated by Jordan-Wigner transformation followed by bosonization of the fermionic modes~\cite{giamarchibook,nersesyanbook}. We start with the following set of definitions
\beq
    S_j^+ = {\rm e}^{+{\rm i}\phi_j}c_j^\dagger,\qquad
    S_j^- = {\rm e}^{-{\rm i}\phi_j}c_j,\qquad
    S_j^z = n_j - 1/2,
\eeq
where $\phi_j=\pi\sum_{k<j}S^+_kS^-_k$ is the Jordan-Wigner string. A term $h(j) = J_j {\mathbf S}_j\cdot\left({\mathbf S}_{j+1}\times{\mathbf S}_{j+2}\right)$ of three consecutive spins with coupling $J_j$ will transform to ${h(j) \rightarrow J_j \left(T(j) + V(j)\right)}$. The kinetic and interacting terms read
\beqa
    T(j) &=& -\frac{\rm i}4\left(c^\dag_j c^{\vphantom\dag}_{j+1} - c^\dag_jc^{\vphantom\dag}_{j+2} + c^\dag_{j+1}c^{\vphantom\dag}_{j+2}\right) + {\rm h.c.}, \nonumber \\
    V(j) &=& +\frac{\rm i}2\left(c^\dag_jc^{\vphantom\dag}_{j+1}n_{j+2} + n_j c^\dag_{j+1}c^{\vphantom\dag}_{j+2}\right) + {\rm h.c.},
\eeqa
with density operator $n_j=c^\dag_j c^{\vphantom\dag}_j$.
\begin{figure*}
    \includegraphics[width=0.246\textwidth]{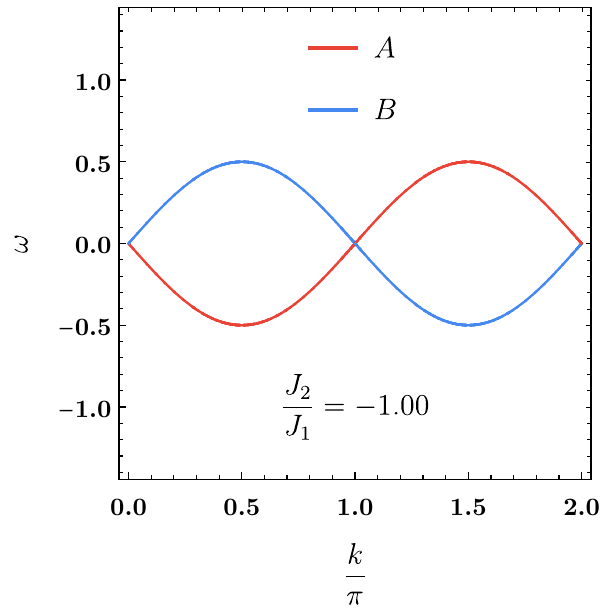}
    \includegraphics[width=0.246\textwidth]{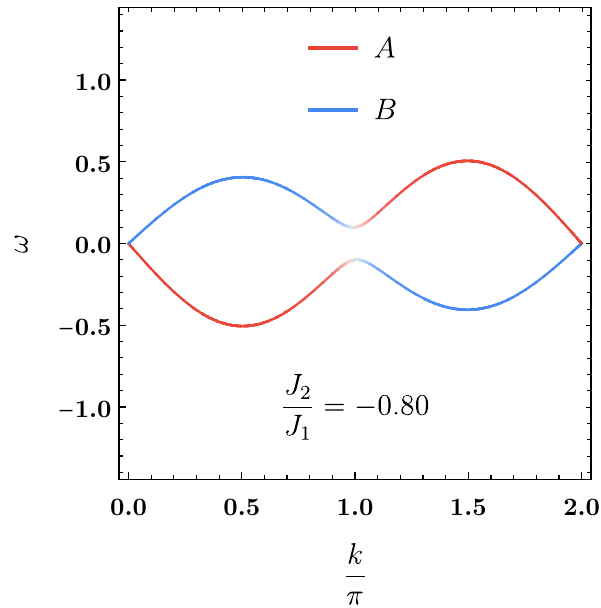}
    \includegraphics[width=0.246\textwidth]{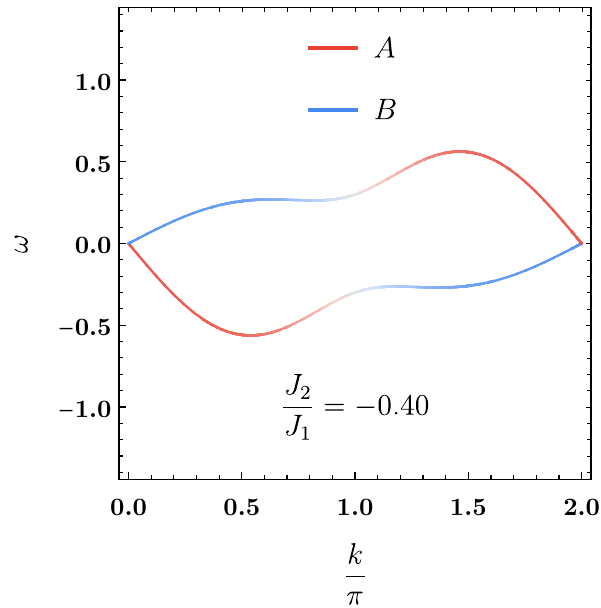}
    \includegraphics[width=0.246\textwidth]{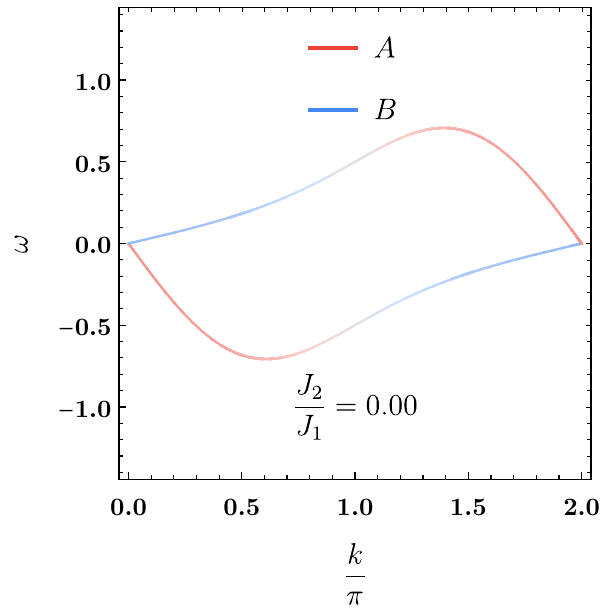}
    \\
    \includegraphics[width=0.246\textwidth]{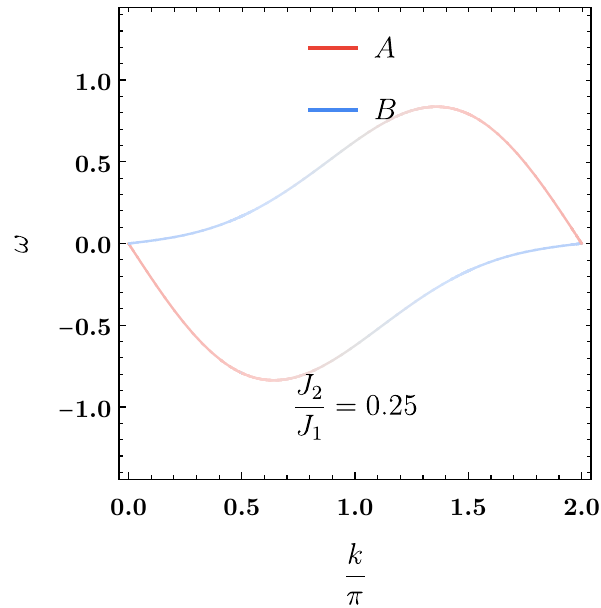}
    \includegraphics[width=0.246\textwidth]{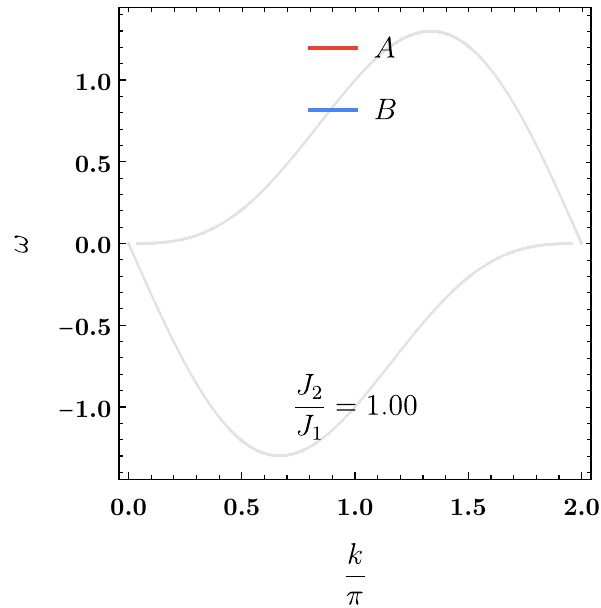}
    \includegraphics[width=0.246\textwidth]{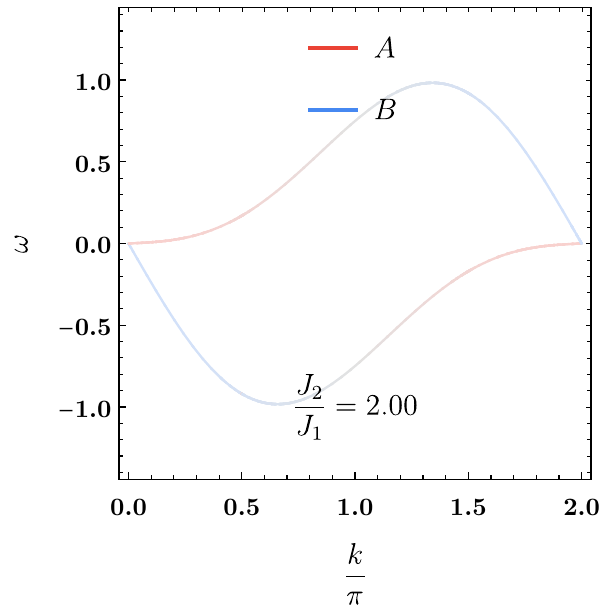}
    \includegraphics[width=0.246\textwidth]{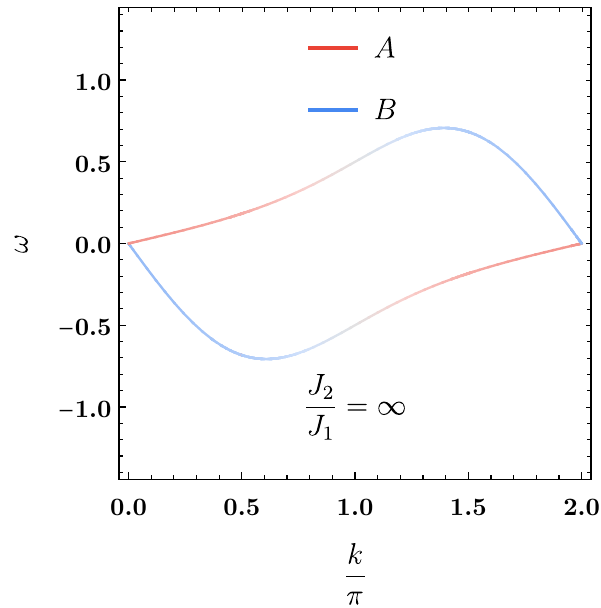}
    \caption{Dispersion for different coupling strengths. The color indicates the polarization of the bands. In the case we study in the paper, both bands are fully polarized and the kinetic dispersion is $\omega(k)=\pm\frac12\sin(k)$. If we tune $\frac{J_2}{J_1}>-1$, we allow for a mixing between $A$ and $B$, and only one of the avoided crossings is preserved.}
    \label{disp}
\end{figure*}
The full kinetic part of the two-site invariant Hamiltonian in Eq.~(\ref{chiralH}) reads
\beq
    H_{\rm kin} = \sum_{j\,{\rm odd}} J_1 T(j) + J_2 T(j+1),
\eeq
which is simply a tight binding model of spinless fermions on a triangular lattice with nonzero fluxes for generic couplings $J_1$ and $J_2$. For the case $J_1 = -J_2$, two parts of the tight-binding Hamiltonian become fully disconnected. To see this, we use a two-site unit cell according to the drawings in the introductory chapter with sublattice $A$ (upper chain) and $B$ (lower chain) and a spinor $d_k=(c_{A,k}\ c_{B,k})^\top$. The tight-binding Hamiltonian in this basis and after a subsequent Fourier transformation reads $H_{\rm kin} = \sum_k d^\dag_k h(k) d^{\vphantom\dag}_k$ and reduces to a sum of $2 \times 2$ matrices given by
\beq
    h(k) = -\frac14 \begin{pmatrix} 2J_1\sin k & {\rm i}J_{+}(1-{\rm e}^{-{\rm i}k}) \\ -{\rm i}J_{+}(1-{\rm e}^{+{\rm i}k}) & 2J_2\sin k \end{pmatrix},
\eeq
where $J_{+} = J_1 + J_2$. Without further restrictions, we assume $J_1 \geq 0$ to avoid unnecessary ambiguities in the ordering of the bands and show the resulting dispersion in Fig.~\ref{disp}. If we consider $J_1 = +1$ and $J_2 = -1$ (i.e. $H_1$), we see two cosine bands which are shifted by $\pm\frac\pi2$ according to the nonzero flux threading the two sublattices. The two Fermi-points correspond to a central charge of $c = 2$. If we slightly increase $J_2$, we find $H(k=0)=0$, that is, there are no scattering processes and the dispersion (i.e., the band crossing) will be left untouched. However, at momentum $k=\pi$, there are strong inter-chain transitions, i.e. $H(k=\pi)=\frac12(J_1+J_2)\sigma_y$ which yields the observed avoided crossing.\\
The interacting part of the Hamiltonian reads
\begin{align}
	H_{\rm int} = \sum_{j\,{\rm odd}} J_1 V(j) + J_2 V(j+1),
\end{align}
which can again be written in terms of the two-site basis
\begin{align}
    \begin{split}
    	H_{\rm int}
    	&= \frac i 2 \sum_j \left[\left(J_1 n_{j+1,A} + J_2 n_{j-1,B}\right)c^\dag_{j,A}c_{j,B} \right. \\
    	&+ \left. \left(J_1n_{j,A} + J_2n_{j+1,B}\right)c^\dag_{j,B}c_{j+1,A} \right] + {\rm h.c.}
    \end{split}
    \label{eq:InteractingHamiltonian}
\end{align}
For better understanding, we visualize all hopping terms of the full Hamiltonian.
\begin{widetext}
	\begin{align*}
		\includegraphics{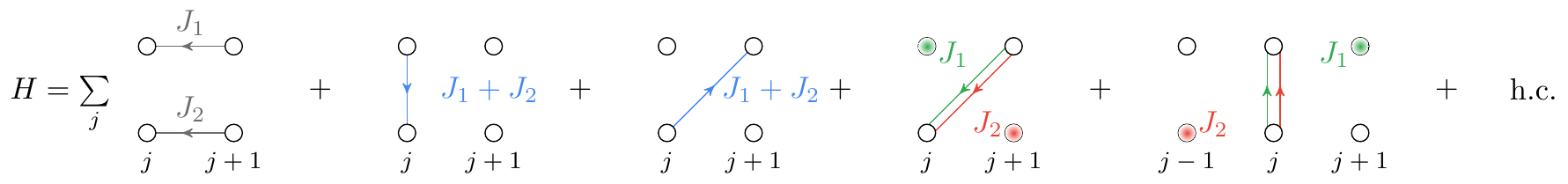}
	\end{align*}
\end{widetext}
Here, the colored sites correspond to the density and the arrows indicate a hopping operator. It is now apparent that the interactions enable the same inter-chain tunnelings as the single particle ones (up to a phase and additional density dependencies). Albeit the isolated study of chiral interactions in existing materials is quite unrealistic, we showed here that it is equivalent to a simple fermionic tight binding model combined with density-assisted hoppings in a quasi one-dimensional ladder setup. A combination of such terms was proposed for experiments in the framework of ultra cold atoms trapped in optical lattices~\cite{densityassistedhoppingTheory} and has been realized very recently~\cite{densityassistedhoppingExp}.\\
In order to understand the gapping mechanism in the model we start from Hamiltonian $H_1$ (${J_1 = -J_2 = +1}$) with two identical bands which are displaced by a phase $\pi$ in momentum space. If we fix the density at a Fermi energy $E_F$ such that the Fermi momentum $k_F$ is in the vicinity of the linear regime of the dispersion, we are allowed to linearize the spectrum. The linearization of such a dispersion is fairly standard and can be described by a Luttinger liquid (LL)
\begin{align}
    \begin{split}
    	H_{\rm LL} = v_F \sum_{\alpha\in\{A,B\}}\sum_k &\left( R^\dag_\alpha(k) (k-k_2^\alpha) R^{\vphantom\dag}_\alpha(k) \right.\\[-0.4cm]
    	& \left. - L^\dag_\alpha(k) (k+k_1^\alpha) L^{\vphantom\dag}_\alpha(k) \right)
    \end{split}
\end{align}
where $R/L_\alpha(k)$ denote right- and left moving modes of the linearized dispersion in the vicinity of the Fermi momenta $k_i^\alpha$ and ${v_F=\frac12\cos(k_2^\alpha)}$ is the Fermi velocity.

We now proceed by rewriting the fermionic modes in terms of right and left moving fields $R/L_\alpha(x)$.
\begin{align}
    \begin{split}
    	c_{A}(x) &\propto \re^{\ri k_1^A x} L_A(x) + \re^{\ri k_2^A x} R_A(x) \\
    	c_{B}(x) &\propto \re^{\ri k_1^B x} L_B(x) + \re^{\ri k_2^B x} R_B(x)\,.
    \end{split}
\end{align}
\begin{figure}[ht]
    \centering
    \includegraphics[width=\columnwidth]{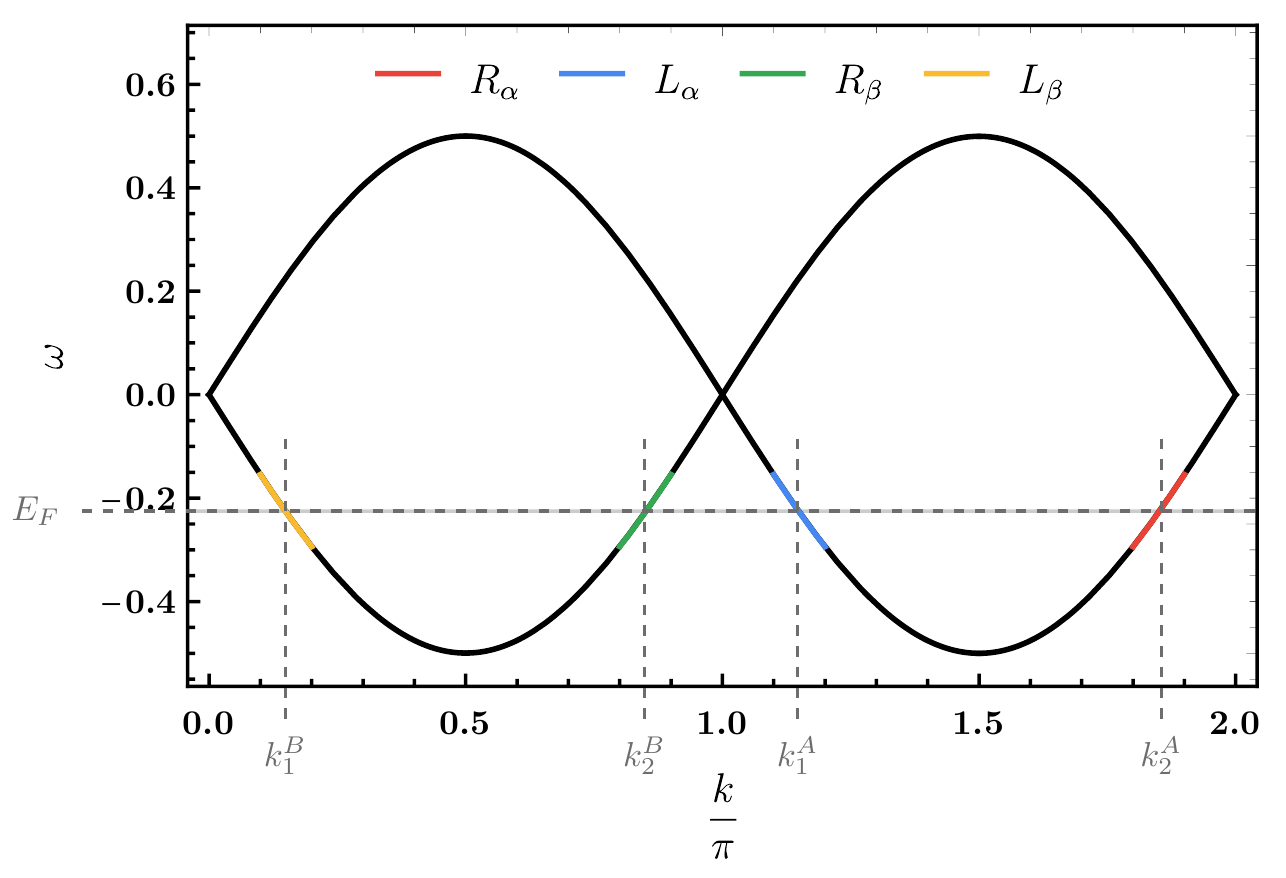}
    \caption{At the Fermi energy $E_F$ the linearized spectrum consists of left and right moving species for each flavor $A,B$.}
    \label{fig:relevant_modes}
\end{figure}
The local densities for both species ${\alpha\in\{A,B\}}$ can be expressed in terms of the new modes
\begin{align}
    \begin{split}
    	n_\alpha(x) &\propto n_{\alpha,R}(x) + n_{\alpha,L}(x) \\
    	&\quad+ \left(\re^{\ri (k^\alpha_1 - k^\alpha_2)x}R^\dag_\alpha(x)L^{\vphantom\dag}_\alpha(x) + {\rm h.c.}\right)\,.
    \end{split}
\end{align}
The gap opening in Fig.~\ref{disp} can be explained by a perturbative expansion in $J_+$. Going back to the general noninteracting Hamiltonian, the $A,B$ scattering processes become
\begin{align}
    J_+\Delta h(x) &\propto \frac{J_+}{2\ri} \left[c^\dag_A c_B - c^\dag_A(x+a)c_B\right] + {\rm h.c.} \label{eq:sp_gapping} \\
    &\approx
    \frac{J_+}{\rm 2\ri}\left(2R_A^\dag(x)L_B + \left(\partial_xL_A^\dag(x)\right)R_B\right)+{\rm h.c.}\ . \nonumber
\end{align}
The first term ($\propto R^\dag_A L_B$) is a relevant, the other an irrelevant perturbation to the Luttinger liquid Hamiltonian.\\
Among many other (supposably irrelevant) terms of the interacting Hamiltonian in Eq.~\ref{eq:InteractingHamiltonian} which we do not devote our main focus to, we obtain a marginal term
\begin{align}
    H_{\rm int} \approx -2\sum_j n_{AB}(j) \Delta h(j)\,.
\end{align}
This yields up to a sign the same scattering process between the sublattices as encountered in Eq.~\ref{eq:sp_gapping}, only here it is multiplied by the density
\begin{align}
    \begin{split}
    	n_{AB}(j) &= J_1\left(n_{A,L}(j) + n_{A,R}(j)\right) \\
    	&\quad+ J_2\left(n_{B,L}(j) + n_{B,R}(j)\right)\,.
    \end{split}
\end{align}
The simple Abelian bosonization approach presented in this chapter is already sufficient to explain two important features. First, by coupling right movers of sublattice $A$ with left movers of sublattice $B$ a gap opens up around $k=\pi$. This removes one of the two Fermi points and the interaction drives the system from a central charge $c=2$ to $c=1$ in a similar fashion as the non-interacting term. Furthermore, it explains the chiral current propagation of the low-lying energy excitations observed in the exact diagonalization, see Fig.~\ref{figpat1}. To predict the long-distance behavior of correlation functions, we resort instead to more sophisticated non-Abelian bosonization techniques. In Appendix~\ref{append2}, we recap the continuum limit of the model by Huang and coauthors following Ref.~\cite{continuum1} and present the expected analytic behaviour for spin-spin correlations.

\section{Methods}
\label{sec3}

The numerical method that we used to compute the ground state properties of the model is infinite-DMRG~\cite{iDMRG}. The iDMRG algorithm has been extensively discussed many times in the literature, see, e.g., Sec.~III of Ref.~\cite{romankai}. In our specific implementation we used the 2-site update for an infinite MPS with a 2-site unit cell. Each physical index of the MPS has dimension 4, and describes the upper and lower spin-1/2 of each rung of the ladder. Moreover, we implemented $SU(2)$ symmetry using the scheme that we described in Ref.~\cite{ourSU2}, based on the formalism of fusion trees to target the $SU(2)$-symmetric ground state of the ladder configuration $J_1 = -J_2 = -1$ ($H_1$). Therefore, the physical index of the MPS carries the quantum numbers $\frac 1 2 \otimes \frac 1 2 = 0 \oplus 1$.

Apart from the implementation of $SU(2)$ symmetry, our iDMRG method heavily relies on the correct implementation of the MPO for the Hamiltonian with $SU(2)$ symmetry. While the case of 3-spin interactions is implicitly discussed in the literature, a more in-depth discussion of this case would be particularly useful. We describe the details of how to implement such an MPO in Appendix\ref{append3}.

\section{Results}
\label{sec4}

\subsection{Energy convergence}

\begin{figure}
	\centering
	\includegraphics[width=\columnwidth]{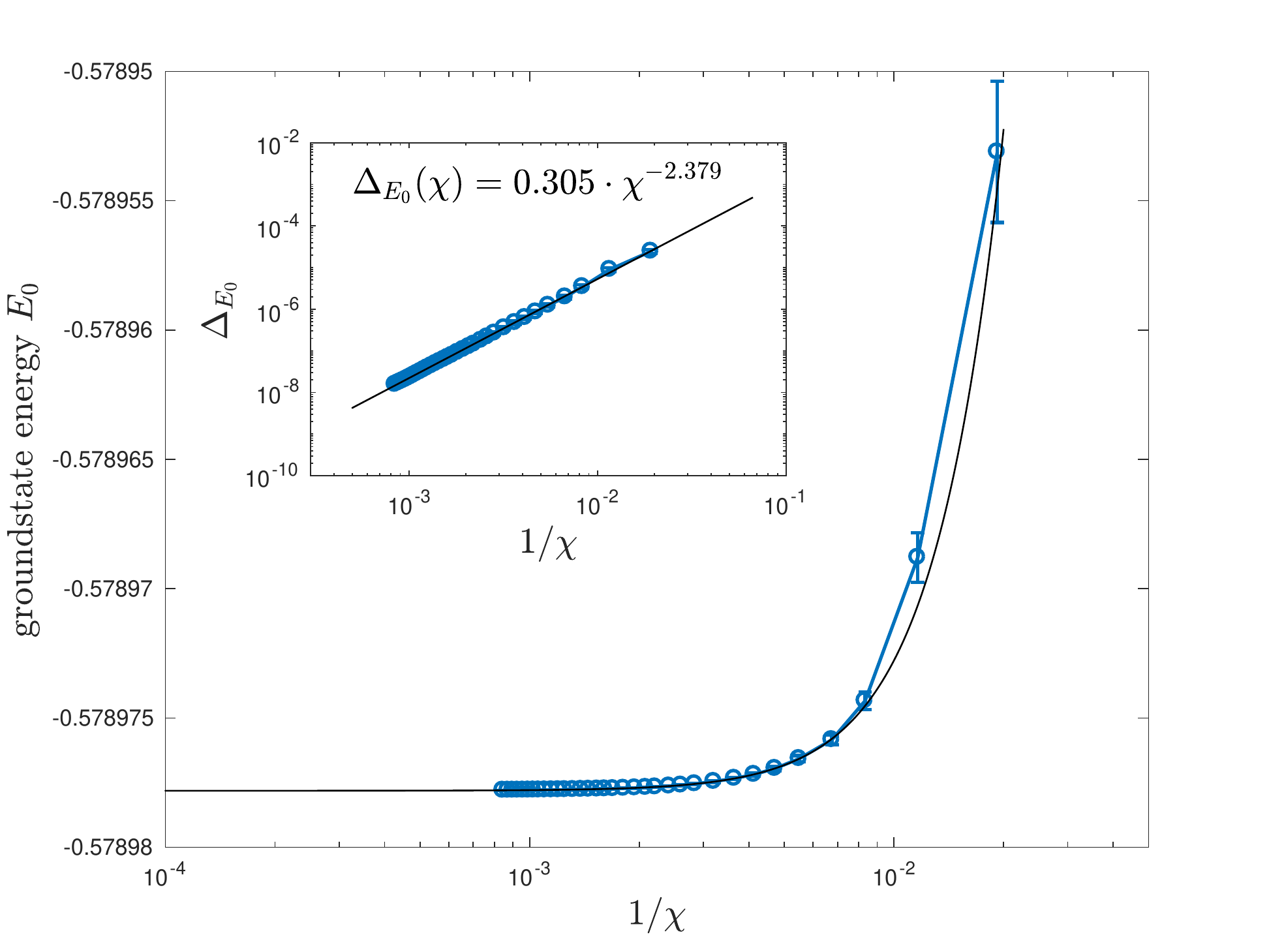}
	\caption{Convergence of the ground state energy $E_0$ with the MPS bond dimension $\chi$. Here we use the discarded weight (see also Fig.~\ref{ener2}) as an estimate of the relative error for the ground state energy in each simulation. The inset shows the convergence of the error $\Delta_{E_0} = E_0 - E_0 (\chi \rightarrow \infty)$ between the energy and its extrapolated value.}
	\label{ener1}
\end{figure}

\begin{figure}
	\centering
	\includegraphics[width=\columnwidth]{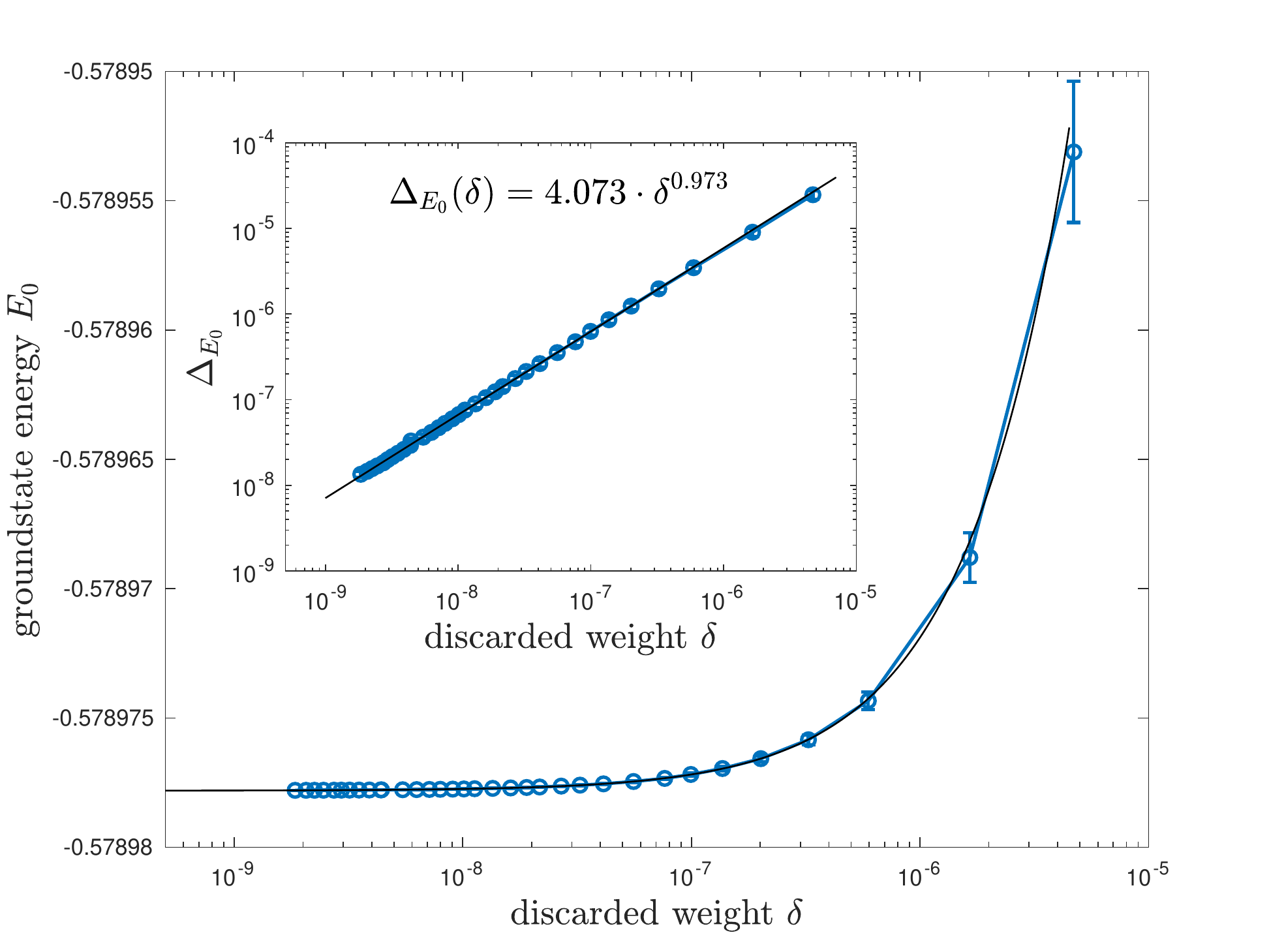}
	\caption{Convergence of the ground state energy $E_0$ with the iDMRG discarded weight. The inset shows the convergence of the error $\Delta_{E_0} = E_0 - E_0 (\chi \rightarrow \infty)$ between the energy and its extrapolated value.}
	\label{ener2}
\end{figure}

Let us start by showing the results for the energy convergence of our $SU(2)$ iDMRG code for the considered chiral Hamiltonian on a ladder. In order to give an overview of simulation parameters and the corresponding irreps on the virtual bonds, we listed some examples in Table~\ref{tab:IrrepsVirtualBondsMPS}. The convergence of the energy with the effective MPS bond dimension $\chi$ is shown in Fig.~\ref{ener1}, where we use up to $\chi \approx 1200$.
In the aforementioned figure, convergence is only approached for very large bond dimension $\chi\gtrsim1000$, which would be difficult to reach without $SU(2)$ symmetric tensors.
In Fig.~\ref{ener2} we show a similar plot, namely, the convergence of the ground state energy with the discarded weight in the iDMRG approximation. As is well known, this allows for a better extrapolation to the $\chi \rightarrow \infty$ limit \cite{extrapol}. In our case, we obtain an estimate of
\begin{align}
	E_0 (\chi \rightarrow \infty) \approx \num{-0.578978 \pm 0.000002}.
\end{align}

\begin{table}
	\centering
	\begin{tabular}{||c|c|c||}
	\hline
	~~$\chi_{\rm sym}$~~	& ~~~$\chi$~~~ 	& ~~~~~~~~virtual bond irreps~~~~~~~ \\
	\hline
	\hline
		50					& 148		& $0_{14} \oplus 1_{24} \oplus 2_{11} \oplus 3_{1}$\\
		100					& 312		& $0_{26} \oplus 1_{46} \oplus 2_{24} \oplus 3_{1}$\\
		150					& 480		& $0_{37} \oplus 1_{68} \oplus 2_{38} \oplus 3_{7}$\\
		200					& 652		& $0_{48} \oplus 1_{89} \oplus 2_{52} \oplus 3_{11}$\\
		250					& 834		& $0_{58} \oplus 1_{110} \oplus 2_{65} \oplus 3_{16} \oplus 4_{1}$\\
		300					& 1008		& $0_{69} \oplus 1_{130} \oplus 2_{80} \oplus 3_{20} \oplus 4_{1}$\\
		350					& 1184		& $0_{80} \oplus 1_{149} \oplus 2_{96} \oplus 3_{24} \oplus 4_{1}$\\
		\hline
	\end{tabular}
	\caption{Symmetric and effective bond dimensions for several simulations together with the irreps and degeneracies on the virtual bonds of the MPS.}
	\label{tab:IrrepsVirtualBondsMPS}
\end{table}

In the insets of Fig.~\ref{ener1} and Fig.~\ref{ener2} we plot the convergence of the error $\Delta_{E_0} = E_0 - E_0 (\chi \rightarrow \infty)$ between the energy and its extrapolated value, as a function of the inverse effective MPS bond dimension $1/\chi$ and the discarded weight respectively \footnote{Generally the convergence of the ground state energy with respect to the discarded weight should be fitted with a linear function~\cite{schollwoeck}. This however is in good agreement with our power-law fit that shows an exponent of 0.973.}.

\subsection{Entanglement}

\begin{figure}
	\centering
	\includegraphics[width=\columnwidth]{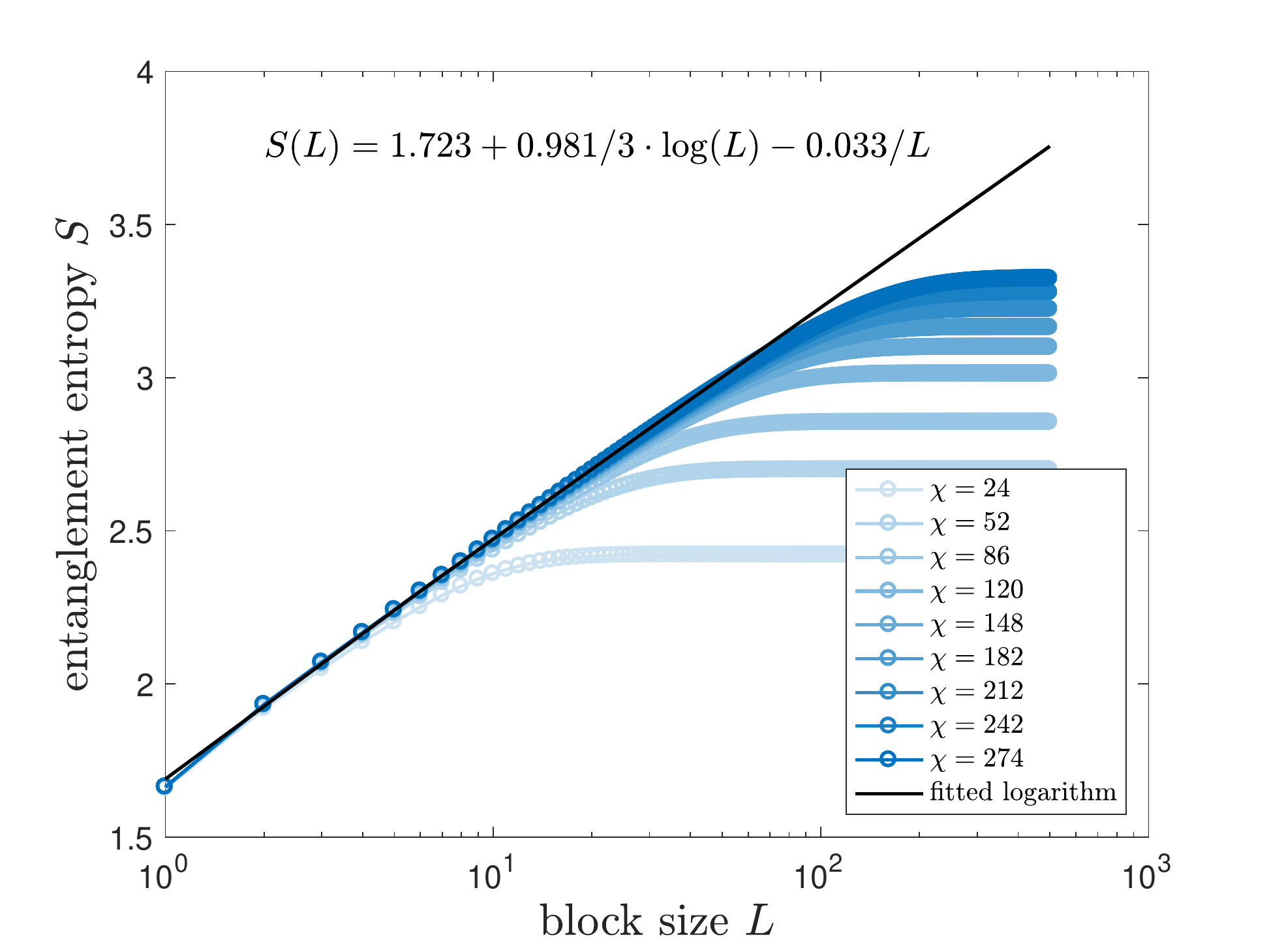}
	\caption{Scaling of the entanglement entropy $S(L)$ for a block of size $L$ and different bond dimensions $\chi$. { Computing the entanglement entropy of a block is restricted to only moderate bond dimensions due to a higher computational cost of $\mathcal O(\chi^5)$ compared to the semi-infinite chain which scales like $\mathcal O(\chi^3)$.}}
	\label{ent1}
\end{figure}

\begin{figure}
	\centering
	\includegraphics[width=\columnwidth]{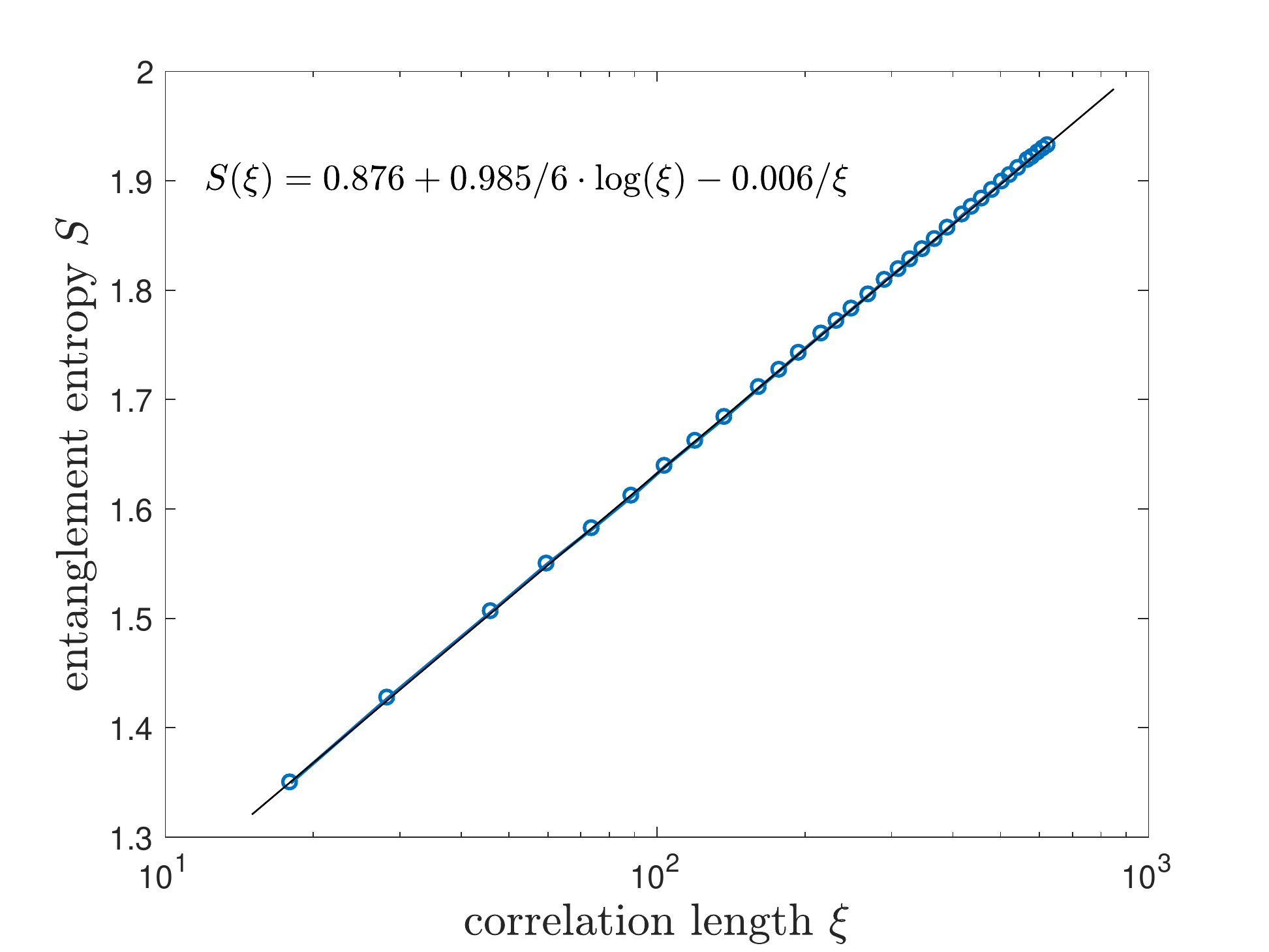}
	\caption{Scaling of the entanglement entropy $S(\xi)$ for half an infinite chain with the MPS correlation length $\xi$ for up $\chi = 1184$.}
	\label{ent2}
\end{figure}

We studied several entanglement figures of merit in our system. First, in Fig.~\ref{ent1} we show the scaling of the entanglement entropy $S(L)$ of a block of length $L$ {with two open ends}, for different values of the bond dimension $\chi$. The computational cost of this calculation is $O(\chi^5)$, as opposed to the $O(\chi^3)$ cost infinite-DMRG. Thus we are restricted to only moderate values of the bond dimension for the calculation of the entanglement entropy. Before reaching saturation due to finite-$\chi$ for large block sizes, the data follows a CFT scaling $ S(L) \sim c / 3 \log L $ \cite{cftscaling}. In our case, this implies that ${c \approx 1}$, as seen in the plot, where we also show our best fit including $O(1/L)$ subleading  corrections. Furthermore, we also analyzed the scaling of the entanglement entropy of half an infinite chain with the MPS correlation length $\xi$, a calculation that instead scales like $O(\chi^3)$. The result shown in Fig.~\ref{ent2} matches perfectly a CFT scaling $S(\xi) \sim c/6 \log \xi$, again with central charge $c \approx 1$.

\begin{figure}
	\centering
	\includegraphics[width=\columnwidth]{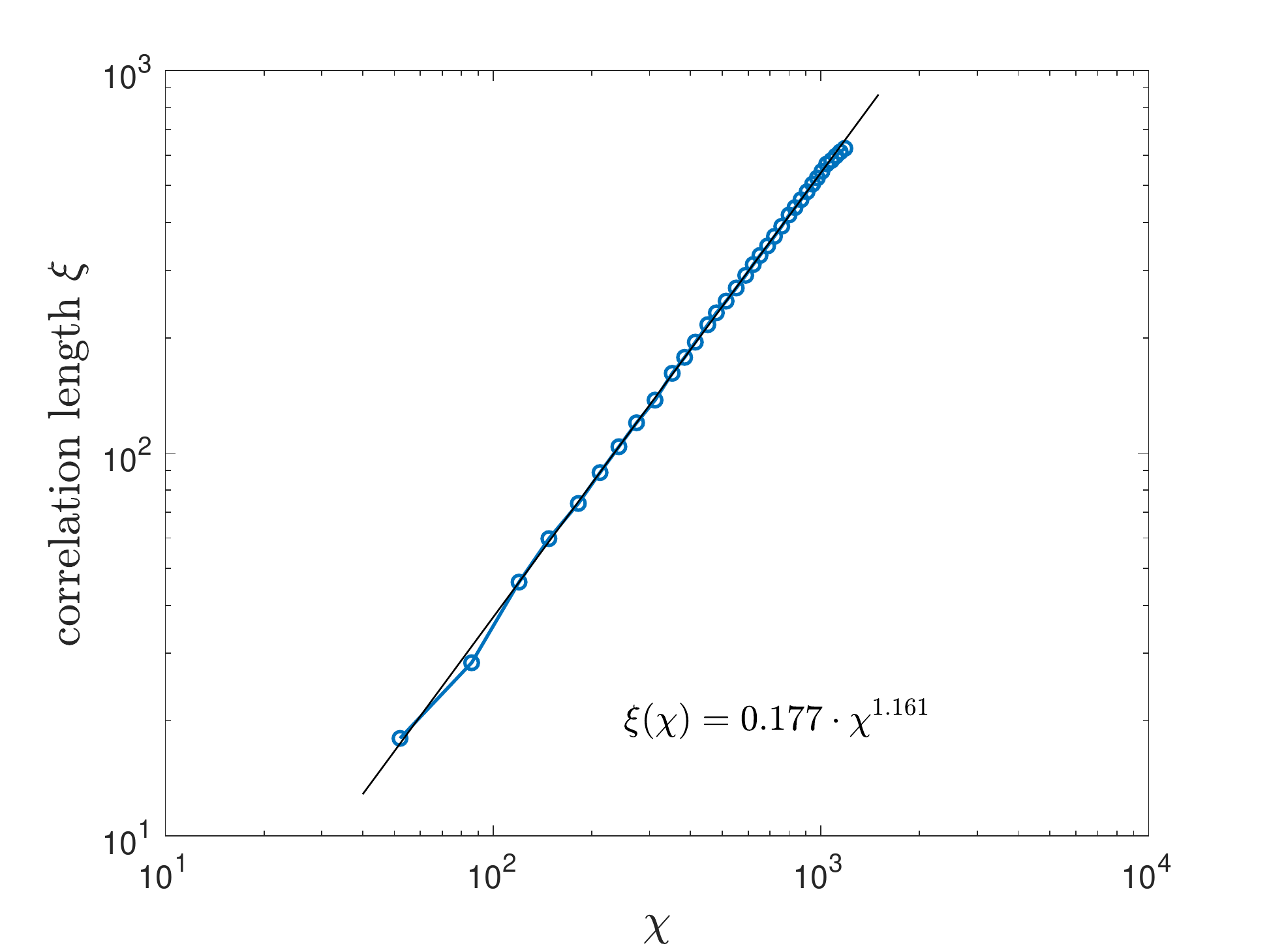}
	\caption{Scaling of the MPS correlation length $\xi$ with the bond dimension $\chi$.}
	\label{ent3}
\end{figure}

In order to assess the consistency of our calculations, we computed the finite-entanglement scaling of the MPS correlation length $\xi$ with the bond dimension $\chi$ \cite{finent}. In Fig.~\ref{ent3} we see that this follows a perfect algebraic fit $\xi \sim \chi^\kappa$, with exponent $\kappa \approx 1.16$, which following the \emph{approximate} formula $\kappa \approx 6/\left(c\left(\sqrt{12/c} + 1 \right)\right)$ \cite{finent}, is again compatible with $c \approx 1$.

\begin{figure}
	\centering
	\includegraphics[width=\columnwidth]{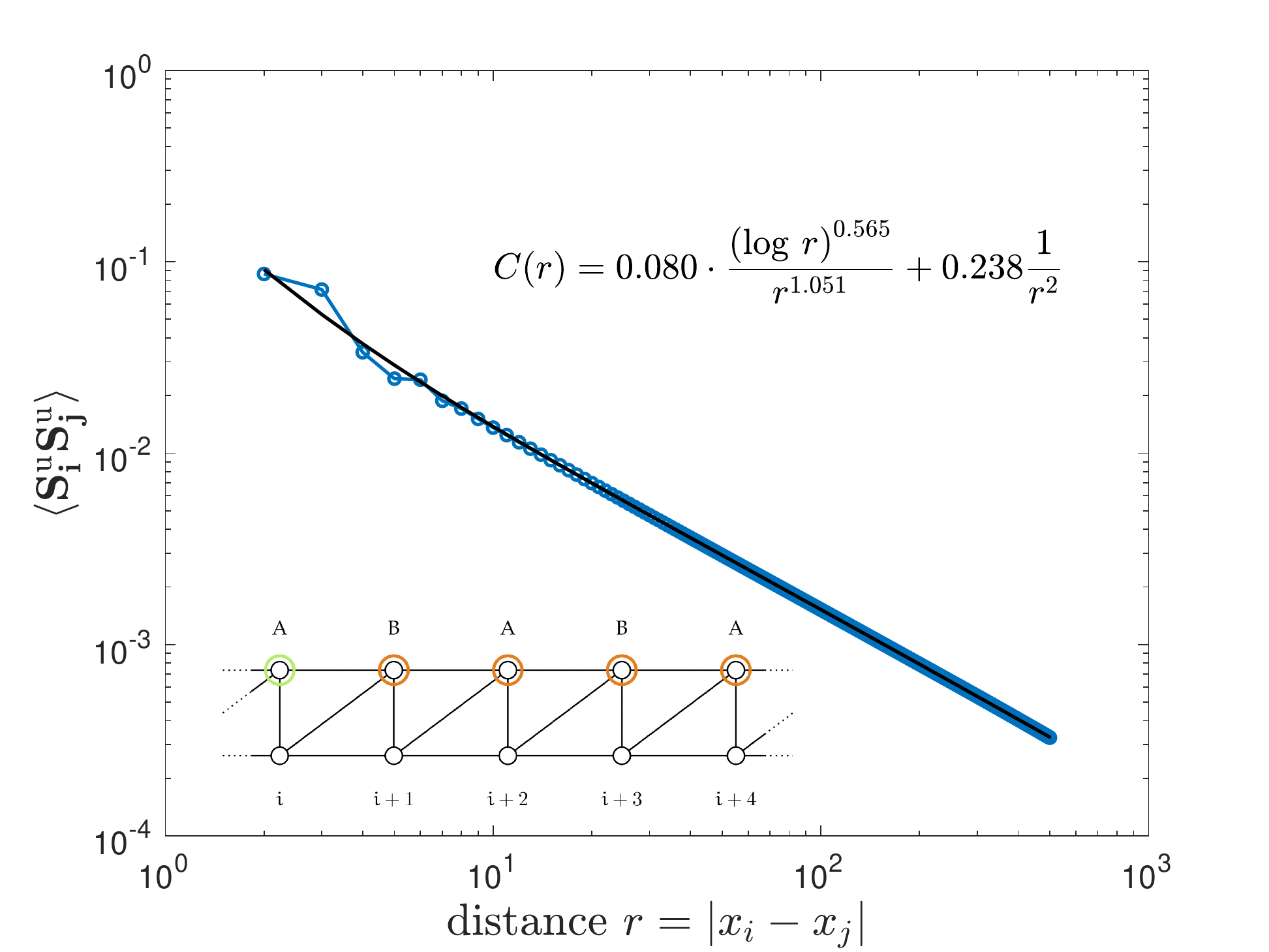}
	\caption{Spin-spin correlation function between spins in the same chain. The two-point correlation function is expected to follow $C(r) \sim (\log\, r)^{1/2} / r$, which is in good agreement with the numerical data~\cite{logcorrections,dimerdimer}. The exponent for the short-distance term is fixed.}
	\label{cr1}
\end{figure}

\subsection{Correlation functions}

In order to assess the criticality of the system, we computed a number of $SU(2)$-invariant correlation functions $C\left(r=\left|i-j\right|\right)$ in the system, with positions $i,j$ and relative distance $r$ (notice that the sites $i,j$ can belong to different ladder legs). We observed algebraic decays and critical exponents, following $C(r) \sim r^{-\alpha}$. We computed all correlation functions for an MPS with bond dimension $\chi = 1184$ ($\chi_\text{sym} = 350$). 

\begin{figure}
	\centering
	\includegraphics[width=\columnwidth]{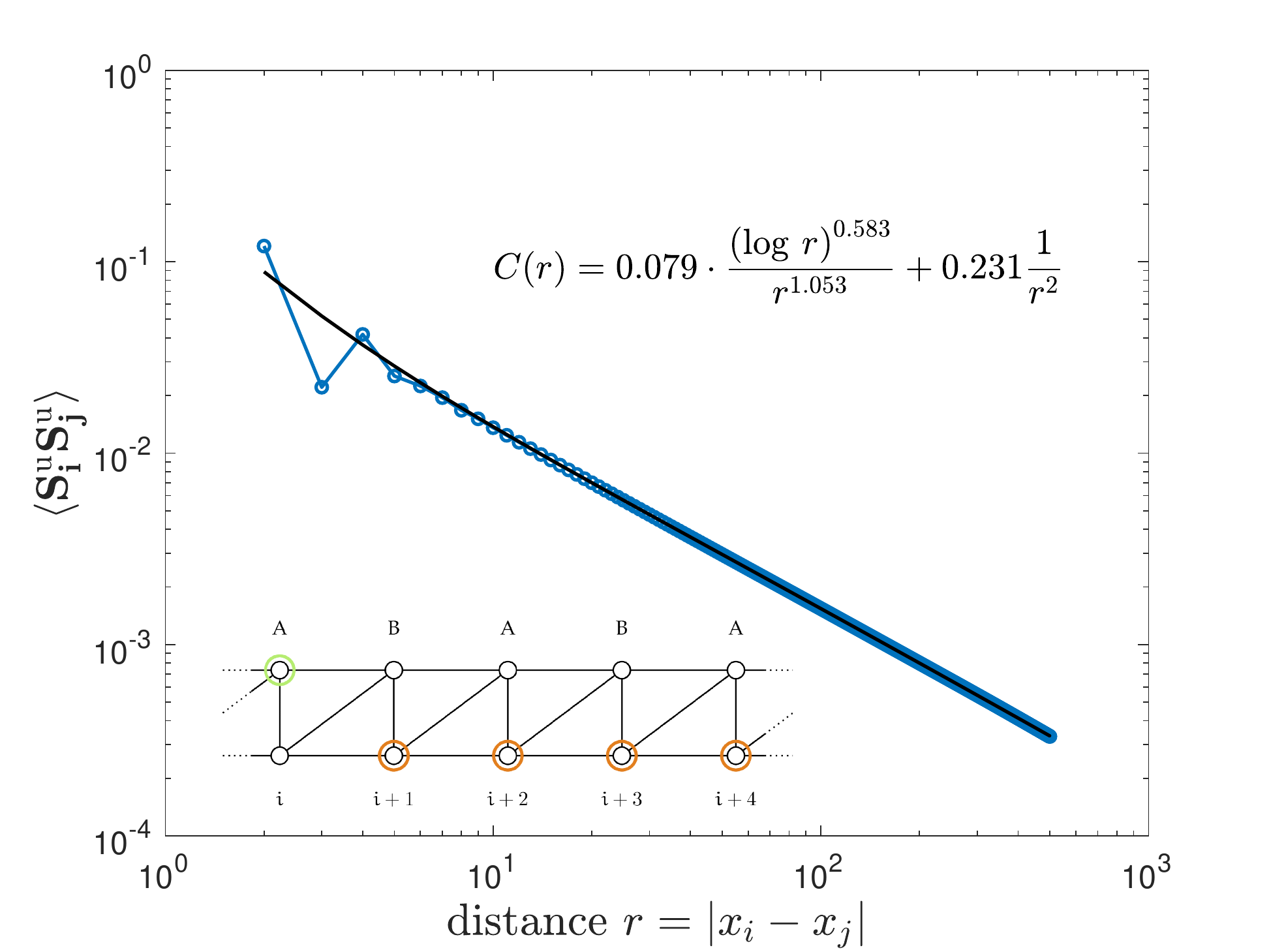}
	\caption{Spin-spin correlation function between spins in different chains. The two-point correlation function is expected to follow $C(r) \sim (\log\, r)^{1/2} / r$, which is in good agreement with the numerical data~\cite{logcorrections,dimerdimer}. The exponent for the short-distance term is fixed.}
	\label{cr2}
\end{figure}

\begin{figure}
	\centering
	\includegraphics[width=\columnwidth]{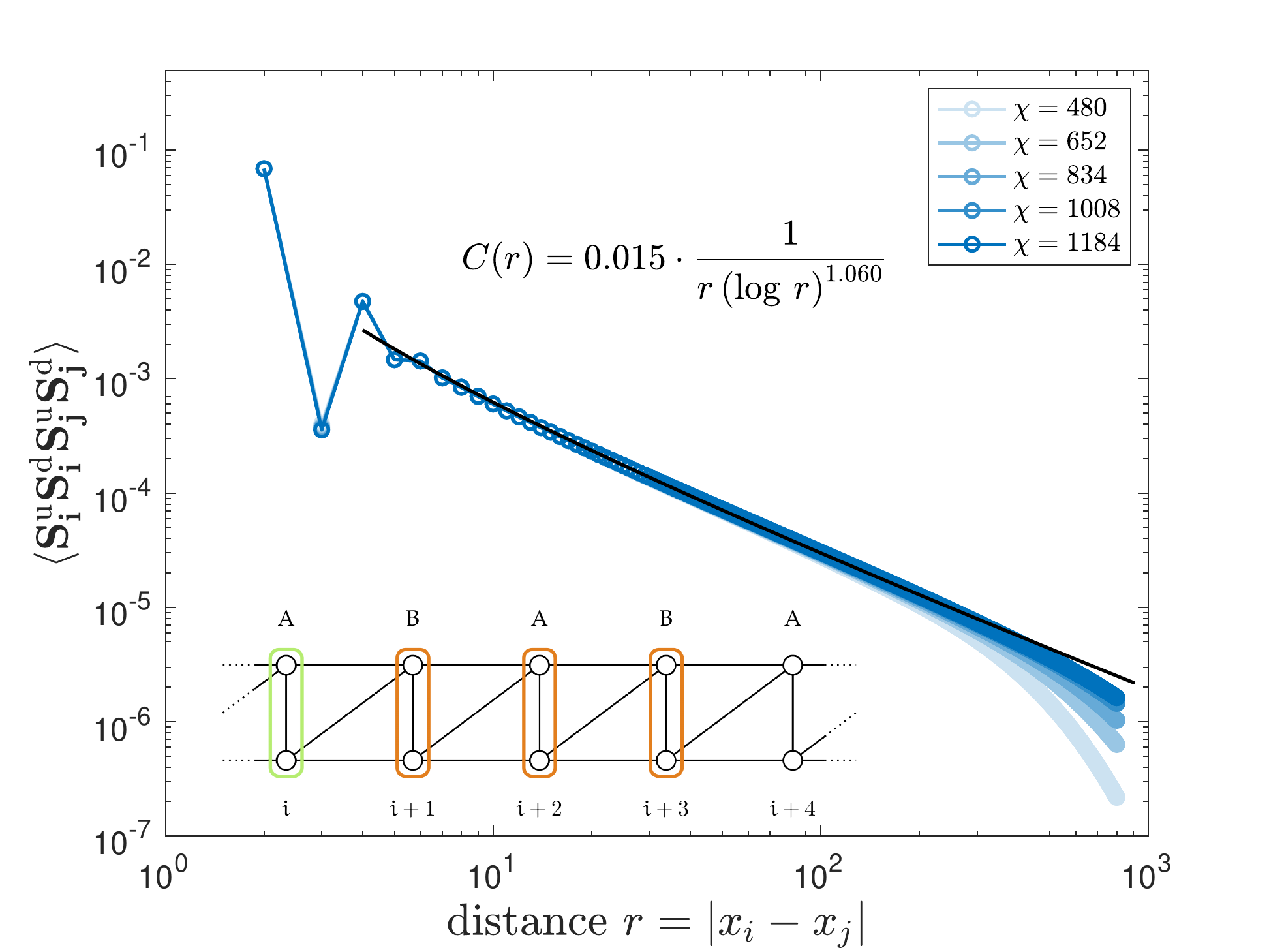}
	\caption{Dimer-dimer correlation function between vertical dimers with logarithmic corrections. The correlations for smaller bond dimensions show an exponential tail due to the final amount of entanglement in the MPS. For larger bond dimension the correlation is expected to follow the fitted function to even larger separation distances.}
	\label{cr5}
\end{figure}

\subsubsection{Spin-spin correlator}

First, we computed the spin-spin correlation functions $\langle {\bf S}_i^u  {\bf S}_j^u \rangle$ and $\langle {\bf S}_i^u  {\bf S}_j^d \rangle$, where $i,j$ are the rung indices and $u,d$ the leg indices, respectively. As shown in Fig.~\ref{cr1} and Fig.~\ref{cr2}, we observe a similar algebraic decay in both cases with exponent $\alpha \approx 1$. This indeed matches the expectation from the continuum limit calculation in Eq.~(\ref{dec}). Additional logarithmic corrections for large distances are conform with the analytic prediction~\cite{logcorrections,dimerdimer}.

\subsubsection{Dimer-dimer correlator}

Next, we studied the dimer-dimer correlation function between vertical dimers $\langle ({\bf S}_i^u  {\bf S}_{i}^d) ({\bf S}_j^u  {\bf S}_{j}^d) \rangle$. The four-body correlation is corrected by all possible disconnected parts, namely $\langle {\bf S}_i^u  {\bf S}_{i}^d \rangle \langle {\bf S}_j^u  {\bf S}_{j}^d \rangle$, $\langle {\bf S}_i^u  {\bf S}_{j}^u \rangle \langle {\bf S}_i^d  {\bf S}_{j}^d \rangle$ and $\langle {\bf S}_i^u  {\bf S}_{j}^d \rangle \langle {\bf S}_i^d  {\bf S}_{j}^u \rangle$ with appropriate factors. The result is shown in Fig.~\ref{cr5}, where the decay fits very well an algebraic decay, as expected for criticality, with decay exponent $\alpha \approx 5/4$.

\begin{figure}
	\centering
	\includegraphics[width=\columnwidth]{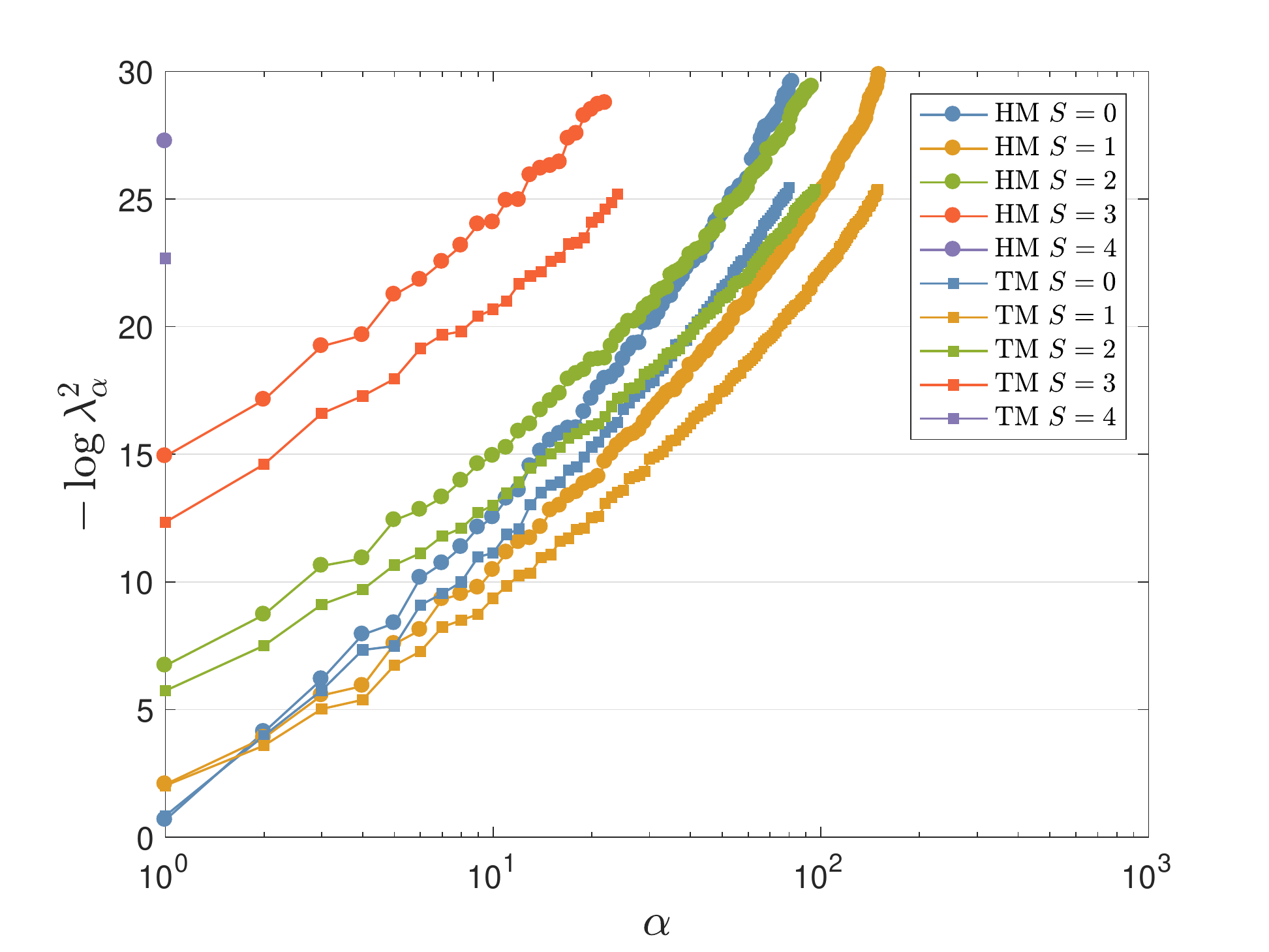}
	\caption{Entanglement spectrum for the triangle ladder model (TM, squares) and the Heisenberg spin chain (HM, dots), with multiplets organized according to their spin sector $S$. Every point is a $(2S+1)$-plet.}
	\label{entanglementSpectrum_1}
\end{figure}

\subsection{Entanglement spectrum}

In order to further characterize the model we also studied the entanglement spectrum of half an infinite chain. The singular values are readily available and their distribution is according to the virtual bond irreps shown in Table~\ref{tab:IrrepsVirtualBondsMPS}, where each spin sector $S$ comes with a $2S+1$ degeneracy. Notice that, by construction, $S$ will always be an integer, because of the coarse-graining of the two spin-$1/2$'s from the upper and lower legs to form the MPS.

In Fig.~\ref{entanglementSpectrum_1} we show our results for the entanglement energies $\varepsilon_\alpha \equiv - \log \lambda_\alpha^2$ with $\lambda_\alpha$ the Schmidt coefficients of half an infinite chain. The results are organized according to the different spin sectors $S = 0, 1, 2, 3$ and $4$ of the bond dimension (so that each point in the plot is indeed a $(2S+1)$-plet). Results for our ladder model with triangles (TM) are given by the squares. In addition, we compare the results of the chiral ladder to the values of $\varepsilon_\alpha$ that we obtain when computing the ground-state of the spin-$1/2$ Heisenberg spin chain (HM) with the same numerical method, and after coarse-graining two sites into one (so that the MPS bond dimension also has integer spin sectors $S$). As can be seen in the plot, both spectra show exactly the same features up to an overall rescaling. This is especially true for the lowest part of the entanglement spectrum, i.e., the largest singular values, which are also most accurate. This is an important observation, because it means that the low energy limits of both lattice systems (TM and HM) have quite probably the same boundary CFT \cite{bcft}, and in practice it means that both limits are probably described by the same $(1+1)$-dimensional CFT. Accordingly, this is a strong indication that the CFT for our chiral ladder model is likely the same than for the Heisenberg spin-$1/2$ chain, i.e., an $SU(2)_1$ WZW theory, which indeed would be in agreement with all our previous results as well as with the $SU(2)$ symmetry of the lattice model. Notice, though, that the continuum Hamiltonian in Eq.~(\ref{clim}) is not yet the one of an $SU(2)_1$ effective theory since it is written in terms of current operators for each leg of the ladder. For completeness we also show the convergence of the entanglement spectrum with the MPS bond dimension in Appendix \ref{append4}.

\section{Conclusions}
\label{sec5}

In this paper we have studied a chiral 2-leg ladder with $SU(2)$ symmetry using an $SU(2)$-invariant iDMRG algorithm.
After getting some intuition about the model by Kadanoff coarse-graining, exact diagonalization, and bosonization, we find numerically that the ground state of the system agrees with a CFT with central charge $c \approx 1$, which is also compatible with previous studies of the continuum limit. In particular, we analyzed the scaling of the entanglement entropy of a block and of half an infinite system, as well as finite-entanglement scaling, ground-state energy convergence, entanglement spectrum, and different correlation functions showing algebraic decay at long separation distances. Our results for the entanglement spectrum are compatible with an $SU(2)_1$ WZW theory in the low-energy limit. Moreover, we explained in full detail how to obtain $SU(2)$-invariant MPOs for 3-spin interactions, something that so far had not been discussed in detail in the literature. Our procedure for constructing such MPOs can be generalized to arbitrary $SU(2)$-invariant interactions on 1d and quasi 1d systems.

Our work motivates further investigations along a number of directions. For instance, it would be interesting to dig deeper into the continuum limit of the model. The case of multi-leg and higher-spin ladders could also be analyzed with techniques similar to the ones in this paper. This would be particularly interesting in order to understand how two-dimensional  physics emerges, and how the gapped/gapless nature of the \emph{chiral} system depends on both the spin and the number of legs. An investigation of configuration $H_3$ with $U(1)$-invariant and/or $SU(2)$-covariant MPS methods would also be relevant in order to understand the overall physics of the chiral ladder configurations that we did not consider here. Investigating similar chiral Hamiltonians in Kagome stripes would also be within reach and could lead to interesting physical insights. Finally, we expect that the simulations in this paper will help us to understand the procedure to simulate chiral quantum spin models in two spatial dimensions with $SU(2)$-invariant tensor networks such as Projected Entangled Pair States (PEPS) \cite{peps, ipeps, su2peps}.

{\bf Acknowledgements.-} We acknowledge discussions with M. Burrello, P. van Dongen, A. Feiguin, I. P. McCulloch, F. Pollmann, S. Singh, and very specially with G. Sierra for clarifications and useful suggestions. We also acknowledge DFG funding through GZ OR 381/3-1 and GZ RI 2345/2-1, as well as the MAINZ Graduate School of Excellence.

\appendix

\section{Derivation of the current operator}
\label{append1}

\begin{figure}
	\centering
	\includegraphics[width=\columnwidth]{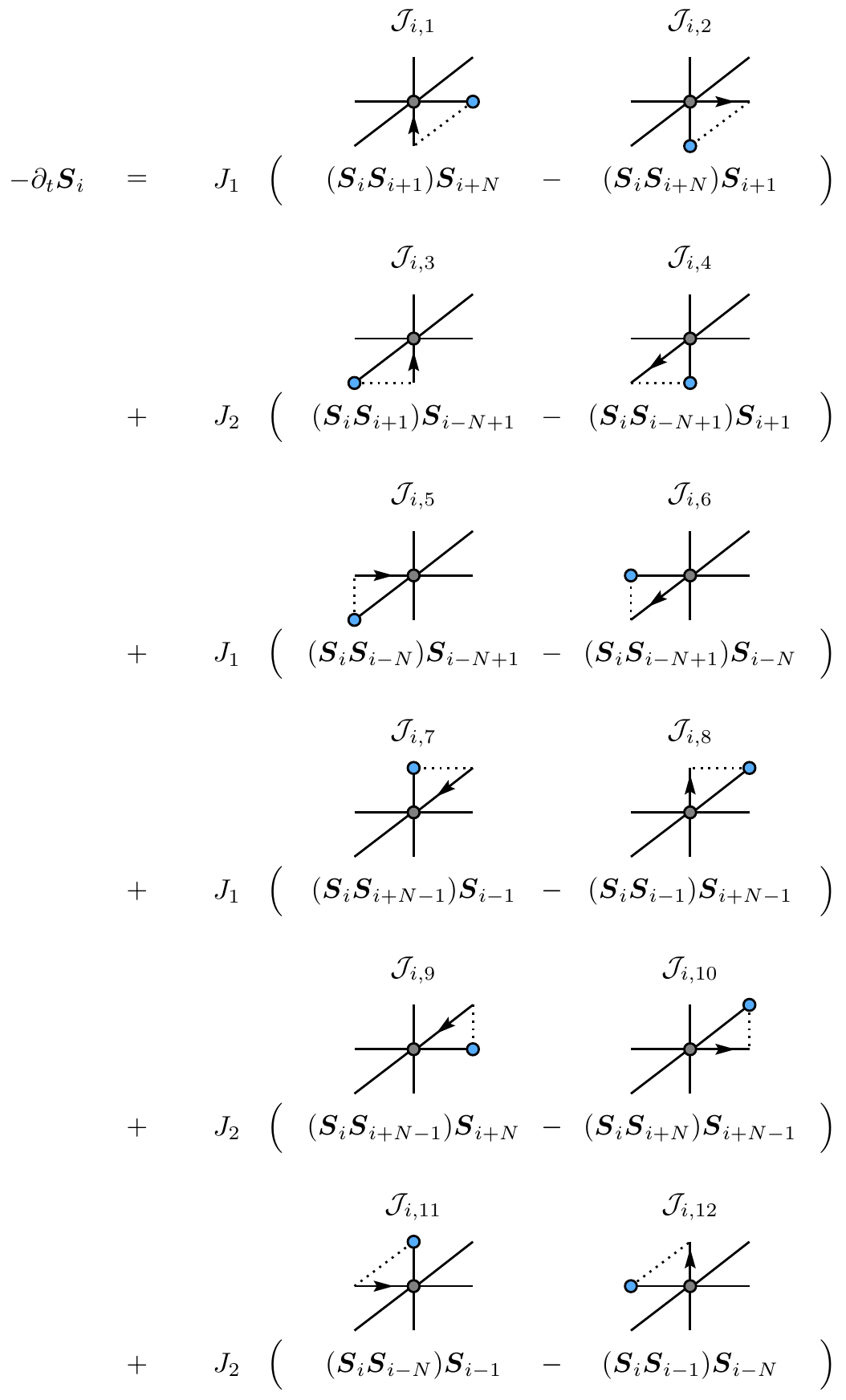}
	\caption{All terms of the spin current operator contributing at each site of the ladder. We define incoming currents as being positive and outgoing currents as being negative. The current in each link of the ladder is a three-site observable which includes a scalar-product (indicated by arrows) between two sites, and the multiplication with $\boldsymbol S$ on the third site (depicted as a blue colored circle).}
	\label{fig:CurrentLinksAll}
\end{figure}

The operators to measure the spin currents flowing in the links of the ladder can be derived from the Hamiltonian
\begin{align}
	H = J_j \epsilon_{\alpha \beta \gamma} S_{j}^{\alpha} S_{j+1}^{\beta} S_{j+2}^{\gamma}
\end{align}
using the Heisenberg equation of motion. Kirchhoff's law is satisfied for all the nodes in the lattice, and we can compute the spin currents passing each node according to
\begin{align}
	\begin{split}
		-\frac{\partial S_i^\mu}{\partial t}
		&= -{\rm i} \Big\lbrack S_i^\mu , H \Big\rbrack = -{\rm i} J_j \epsilon_{\alpha \beta \gamma} \Big\lbrack S_i^\mu , S_{j}^{\alpha} S_{j+1}^{\beta} S_{j+2}^{\gamma} \Big\rbrack \\
		&= J_{i} \left( \boldsymbol S_{i} \boldsymbol S_{i+1} S_{i+2}^\mu - \boldsymbol S_{i} \boldsymbol S_{i+2} S_{i+1}^\mu \right) \\
		&\quad+ J_{i-1} \left( \boldsymbol S_{i} \boldsymbol S_{i+1} S_{i-1}^\mu - \boldsymbol S_{i-1} \boldsymbol S_{i} S_{i+1}^\mu \right) \\
		&\quad+ J_{i-2} \left( \boldsymbol S_{i-2} \boldsymbol S_{i} S_{i-1}^\mu - \boldsymbol S_{i-1} \boldsymbol S_{i} S_{i-2}^\mu \right)\, .
	\end{split}
\end{align}
For a general ladder with $N$ chains there are two currents on every link, one for each triangle the link appears in. The resulting twelve different currents per lattice site can be presented in a more intuitive way in Fig.~\ref{fig:CurrentLinksAll}.
For the $N=2$ leg ladder subject to analysis in the paper there are only six terms for every lattice site $i$, e.g. $\mathcal J_{i,n}$ with $n \in \lbrack 1,2,3,4,5,6 \rbrack$ for the upper chain and ${n \in \lbrack 7,8,9,10,11,12 \rbrack}$ for the lower chain assuming periodic boundary conditions. In case of open boundary conditions, one has to disregard non-existent terms at the edges.

\section{Continuum limit}
\label{append2}

The ladder Hamiltonian in Eq.~(\ref{chiralH}) has been considered in the literature in the presence of extra spin-spin Heisenberg-like interactions \cite{continuum1, continuum2, continuum3}. In these works, the low-energy continuum limit of our purely-chiral lattice model $H_1$ (equivalently $H_2$) has also been computed (see, e.g., Appendix A of Ref.~\cite{continuum1}). Here we sketch briefly the main points of this derivation, and discuss some implications.

The key idea to derive the continuum limit is to use the same formalism as in Ref.~\cite{continuum4} (see also Ref.~\cite{continuum5}) to deal with the continuum limit of the antiferromagnetic Heisenberg quantum spin chain. Starting from a Hubbard-like Hamiltonian for fermions with spin (say, electrons), one considers operator $c_{i,\alpha}$ which annihilates an electron of spin $\alpha$ at site $i$. Spin operators are then written in terms of these fermionic operators as
\beq
{\bf S}_i = \frac{1}{2} c^\dagger_{i,\alpha} {\pmb  \sigma}_{\alpha \beta}  c_{i,\beta},
\eeq
with ${\pmb \sigma}$ a vector of Pauli matrices. As explained in Ref.~\cite{continuum4}, in the continuum limit the fermionic field is expanded around the Fermi points $k \approx \pm \pi/2$
\beq
c_{i,\alpha} \rightarrow \psi_\alpha(x) \sim {\rm e}^{-{\rm i} \pi x /2} \psi_{L,\alpha}(x) + {\rm e}^{{\rm i} \pi x /2} \psi_{R,\alpha}(x),
\eeq
with $\psi_{L/R,\alpha}(x)$ slowly-varying fields on the scale of lattice spacing $a$, which annihilate left/right moving electrons. These chiral fermions can be bosonized as
\beq
\psi_{L/R,\alpha}(x) \sim {\rm e}^{-{\rm i} \sqrt{2 \pi} \varphi_{L/R,\alpha}(x)},
\eeq
with $\varphi_{L/R,\alpha}(x)$ chiral bosons. Introducing charge and spin degrees of freedom as
\beqa
\varphi_{L/R,c}(x) &=& \frac{\varphi_{L/R,\uparrow}(x) + \varphi_{L/R,\downarrow}(x)}{\sqrt{2}}, \nonumber \\
\varphi_{L/R,s}(x) &=& \frac{\varphi_{L/R,\uparrow}(x) - \varphi_{L/R,\downarrow}(x)}{\sqrt{2}},
\eeqa
one can see that at half filling (or more restrictively, for one electron per lattice site), a small Hubbard interaction gaps out the charge mode. This can then be integrated out, and the low-energy properties are then described by spin physics. Moreover, for the Heisenberg model, the $SU(2)$ spin symmetry is independently conserved for fields $\psi_{L/R,\alpha}(x)$ \cite{continuum4} which, following Noether's theorem implies that one can write the conserved (chiral) currents
\beq
{\bf J}_{L/R} = \frac{1}{2} \psi^\dagger_{L/R,\alpha} {\pmb  \sigma}_{\alpha \beta}  \psi_{L/R, \beta}.
\eeq
Thus, in the continuum limit, the lattice spin operators can be written as
\beq
a^{-1} {\bf S}_i \approx \left( {\bf J}_L + {\bf J}_R \right) + \frac{1}{2}(-1)^i  \left( \psi^\dagger_{L,\alpha} {\pmb  \sigma}_{\alpha \beta}  \psi_{R, \beta} + h.c. \right).
\eeq
Bosonizing the fermionic fields in the second term and integrating out the charge boson \cite{continuum4} one arrives at the expression
\beq
a^{-1} {\bf S}_i \approx \left( {\bf J}_L + {\bf J}_R \right) + (-1)^i \Theta ~ {\rm tr}(g_{\rm W}  \cdot {\pmb  \sigma}),
\eeq
with $g_{\rm W}(x)$ the Wess-Zumino-Witten (WZW) field (i.e., an $SU(2)$ matrix) and $\Theta$ a non-universal constant. Matrix $g_{\rm W}(x)$ is explicitely given by
\beq
g_{\rm W}(x) =
\begin{pmatrix}
{\rm e}^{{\rm i} \sqrt{2 \pi} \varphi_s} & {\rm e}^{{\rm i} \sqrt{2 \pi} \bar{\varphi}_s} \\
{\rm e}^{-{\rm i} \sqrt{2 \pi} \bar{\varphi}_s} & {\rm e}^{-{\rm i} \sqrt{2 \pi} {\varphi}_s}
\end{pmatrix},
\eeq
with $\varphi_s \equiv \varphi_{L,s} + \varphi_{R,s} \equiv (\varphi_{L,\uparrow}+\varphi_{R,\uparrow}-\varphi_{L,\downarrow}-\varphi_{R,\downarrow})/\sqrt{2}$ and $\bar{\varphi}_s \equiv \varphi_{L,s} - \varphi_{R,s} \equiv (\varphi_{L,\uparrow}-\varphi_{R,\uparrow}-\varphi_{L,\downarrow}+\varphi_{R,\downarrow})/\sqrt{2}$ following notation from Ref.~\cite{leshouches}. Usually one defines ${{\bf n} \equiv \Theta ~ {\rm tr}(g_{\rm W} \cdot {\pmb  \sigma})}$, which physically amounts to a quantum field for the staggered magnetization. Thus one finally arrives at the usual expression
\beq
a^{-1} {\bf S}_i \approx \left( {\bf J}_L + {\bf J}_R \right) + (-1)^i {\bf n}.
\label{spinop}
\eeq
This equation sets the connection between lattice spin operators and chiral spin-current fields. Thus, for a Hubbard-like system of electrons with exactly one electron per site, charge degrees of freedom are frozen and spin physics emerges entirely in terms of these operators. Importantly for our purposes, in Ref.~\cite{continuum1} it was shown that with this substitution, the continuum limit of our chiral Hamiltonian $H_1$ is given by
\beq
H_1 \approx g \int {{\rm d}x \left(  {\bf J}_{L,1} \cdot {\bf J}_{R,2} -  {\bf J}_{R,1} \cdot {\bf J}_{L,2} \right)},
\label{clim}
\eeq
where $g = 4a/\pi$ with $a$ the lattice spacing, and ${\bf J}_{L/R,1/2}$ is the chiral $L/R$ current  for the upper (1) or lower (2) legs of the ladder. Here, the approximation $\approx$ means that the equivalence holds up to irrelevant local perturbations, and in the limit of lattice spacing going to zero. The calculation arriving to this effective Hamiltonian, which we do not reproduce entirely here, makes use of operator product expansions and analyses the existence of irrelevant terms in the continuum Hamiltonian.  Notice that, as expected from the lattice Hamiltonian, the obtained quantum field theory is odd with respect to time-reversal symmetry (i.e., the exchange $L \leftrightarrow R$). As a remark, let us also mention that, recently, it has been shown that Eq.~(\ref{clim}) can also be written in terms of four Majorana fields for each leg \cite{majo}.

In the CFT language, operators ${\bf J}_{L/R}$ are $SU(2)_1$ Kac-Moody chiral currents, and are the ones entering the $SU(2)_1$ WZW model as low-energy effective field theory of the spin-$1/2$ Heisenberg quantum spin chain \cite{continuum4}. Importantly for our purposes, in Ref.~\cite{continuum3} it was shown that the quantum field theory in Eq.~(\ref{clim}) for the chiral spin ladder is an RG fixed point, i.e., ${\rm d}g/ {\rm d} l$ = 0, with $l$ the RG-flow parameter. In other words, the continuum limit of $H_1$ is a scale-invariant quantum field theory in $(1+1)$ dimensions. In combination with unitarity, a theorem by Zamolodchikov and Polchinski \cite{cft2d} implies that this is indeed a $(1+1)$-dimensional conformal field theory (CFT) \footnote{As far as the authors know, this theorem only exists for two-dimensional CFTs, and its generalization to higher dimensions has proven remarkably hard.}. Therefore, we expect a critical behaviour in the numerical simulations of the lattice Hamiltonian for a purely-chiral ladder in Eq.~(\ref{chiralH}).


At this point it is worth mentioning that, using these results for the continuum limit, we can actually predict some of the expected behavior for the correlation functions that we will compute for the lattice model. In particular, let us consider here the spin-spin correlator $\langle {\bf S}_i  {\bf S}_j \rangle$ for, say, the upper leg. Rewriting the continuum limit of spin operators as in Eq.~(\ref{spinop}), one arrives at a expression for the correlator in terms of ${\bf J}_{L/R}$ and ${\bf n}$ fields. Using the operator product expansion for these fields, one can compute the decay of the asymptotic decay of the correlator. These operator product expansions can be found in, e.g., the Appendix of Ref.~\cite{continuum1}. The relevant non-vanishing ones for our case are
\beqa
J^a_M(x_M) J^b_M(0) &\sim& \frac{1}{(2\pi)^2} \frac{\delta_{ab}/2}{x_M^2} + \frac{i \epsilon_{abc}}{2 \pi} \frac{J^c_R(0)}{x_M} , \nonumber \\
n^a(x)n^b(0) &\sim& \frac{1}{2 \pi^2 a} \frac{\delta_{ab}}{(x_L x_R)^{1/2}} , \nonumber \\
J^a_L(x_L) n^b(0) &\sim& i \frac{\epsilon_{abc} n^c(0) +  \delta_{ab} {\rm tr}(g(0))/2\pi a}{4 \pi x_L},  \nonumber \\
J^a_R(x_R) n^b(0) &\sim& i \frac{ \epsilon_{abc} n^c(0) -  \delta_{ab} {\rm tr}(g(0))/2\pi a}{4 \pi x_R},
\eeqa
with $M = L/R$ and $x_{L/R} = v \tau \pm ix$ holonomic/antiholonomic coordinates (where $\tau$ is imaginary time and $v$ the velocity of the spin mode). Expanding the correlator and computing the vacuum expectation value according to the above expressions, the leading contribution at long distances is given by $\sum_a n^a(x)n^a(0)$, such that we obtain
\beq
\langle {\bf S}_i  {\bf S}_j \rangle \propto \frac{1}{|j - i|}
\label{dec}
\eeq
up to multiplicative and additive constants. We will confirm this asymptotic decay later with our numerical simulations, as well as compute a number of other lattice correlation functions.

\section{Construction of \texorpdfstring{$SU(2)$}{TEXT}-invariant MPOs}
\label{append3}

Here we explain how to construct MPOs for different types of $SU(2)$-invariant interactions based on symmetry considerations only. We start with simple MPOs, such as the one for the Heisenberg quantum spin chain, and then move on to more complex interactions such as the three-spin chiral interactions that we consider in this paper.

The ultimate goal is to write the desired Hamiltonian in an $SU(2)$-invariant form, which implies a decomposition in terms of degeneracy tensors and structural tensors (Clebsch-Gordan coefficients) as described by the Wigner-Eckart theorem. In order to do this, we first consider the Clebsch-Gordan coefficients which are eligible for the MPO and determine afterwards which degeneracy factors are necessary in order to construct the correct Hamiltonian, with the constraint that the Hamiltonian must be an $SU(2)$ scalar.

\begin{figure}
	\centering
	\includegraphics[width=0.55\columnwidth]{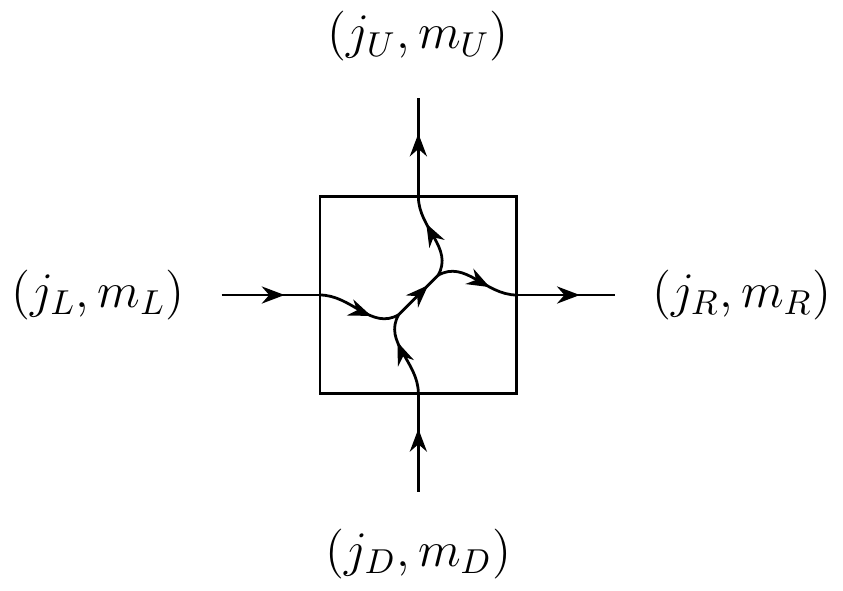}
	\caption{Internal $SU(2)$ structure of a generic MPO tensor with four indices.}
	\label{fig:TikZ_Files_SU(2)_Generic_MPO_LegDefinition_2}
\end{figure}

As a building block, we consider the generic MPO tensor in Fig.~\ref{fig:TikZ_Files_SU(2)_Generic_MPO_LegDefinition_2}, which shows an internal structure due to the presence of four indices in the MPO tensor. The left/right indices are the MPO bond indices, and the up/down indices are the physical indices. Each trivalent tensor corresponds to an intertwiner of $SU(2)$, i.e., a Clebsch-Gordan coefficient. Arrows show the direction of the legs (incoming/outgoing irreps, see Ref.~\cite{ourSU2}).

To show how the construction process works for the MPO, we consider two different Hamiltonians: a 2-site Heisenberg interaction $\mathbf{S} \cdot \mathbf{S}$, and a 3-spin chiral interaction $\mathbf{S} \cdot (\mathbf{S} \times \mathbf{S})$.

\subsection{Heisenberg 2-spin interaction} 
\label{sub:heisenberg_hamiltonian}
\begin{figure}
	\centering
	\includegraphics[width=\columnwidth]{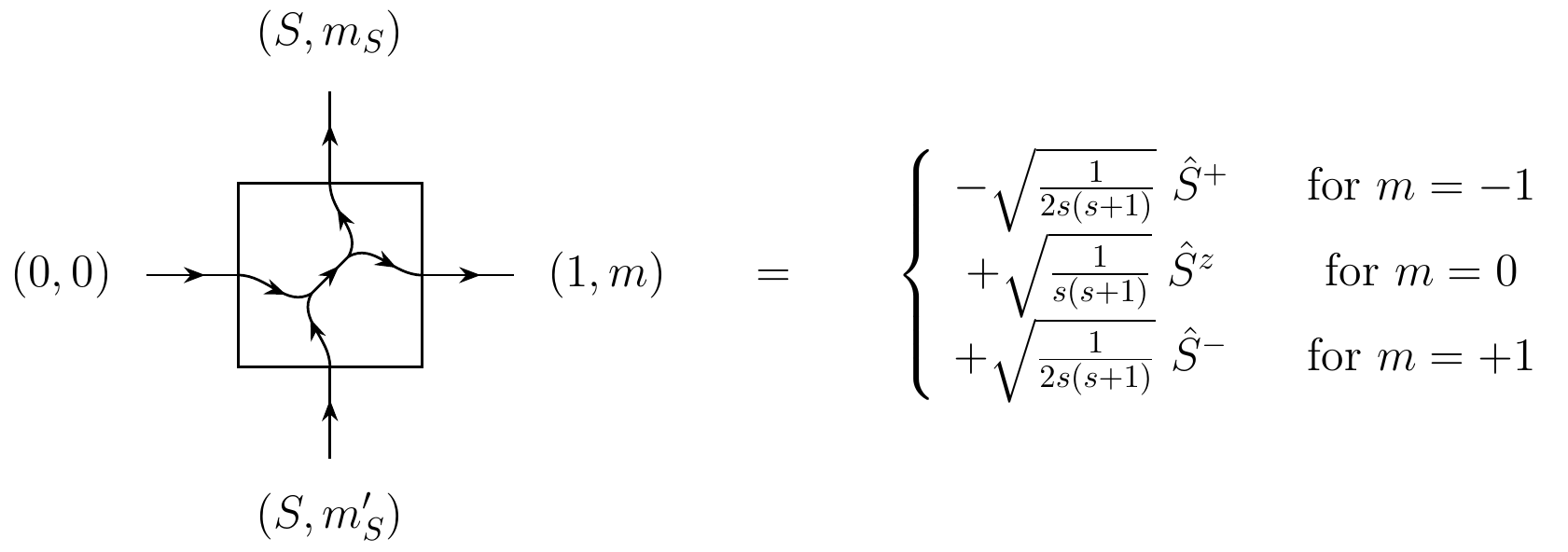}
	\caption{Matrices for the MPO tensor with the left bond index fixed to spin 0 and the right bond index fixed to spin 1, leaving the freedom of choosing $m = -1,0,+1$. Physical indices have spin $S = 1/2$.}
	\label{fig:TikZ_Files_SU(2)_MPO_SdotS_1}
\end{figure}

\begin{figure}
	\centering
	\includegraphics[width=\columnwidth]{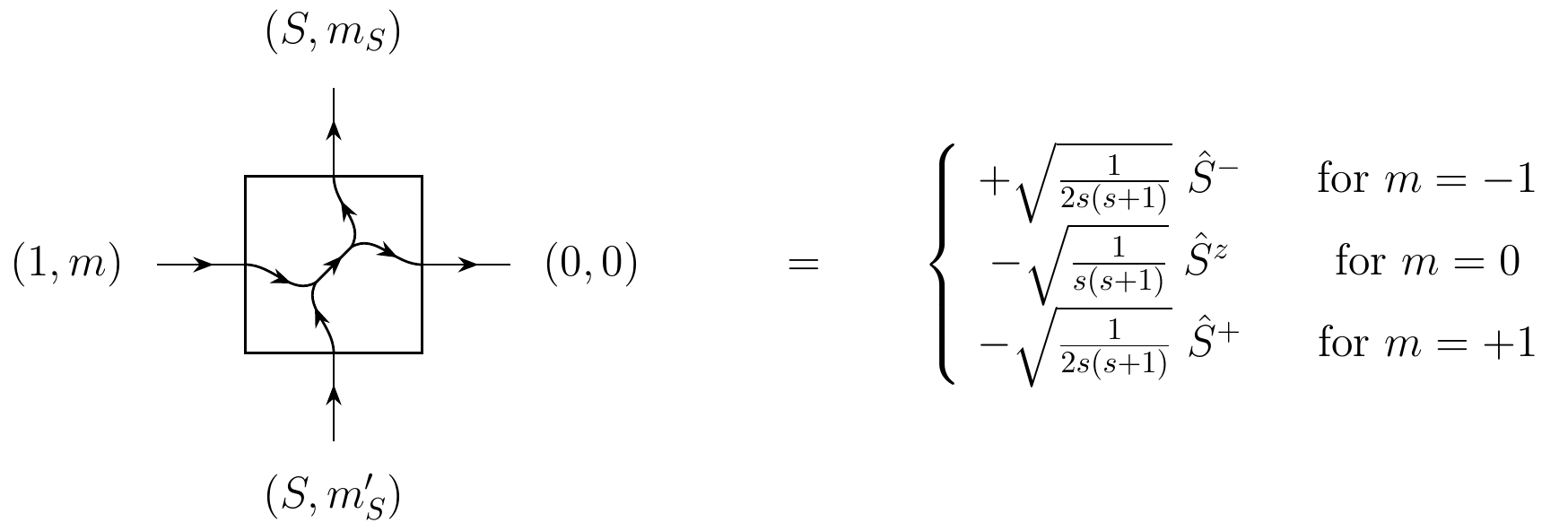}
	\caption{Matrices for the MPO tensor with the left bond index fixed to spin 1 and the right bond index fixed to spin 0, leaving the freedom of choosing $m = -1,0,+1$. Physical indices have spin $S = 1/2$.}
	\label{fig:TikZ_Files_SU(2)_MPO_SdotS_2}
\end{figure}

\begin{figure}
	\centering
	\includegraphics[width=\columnwidth]{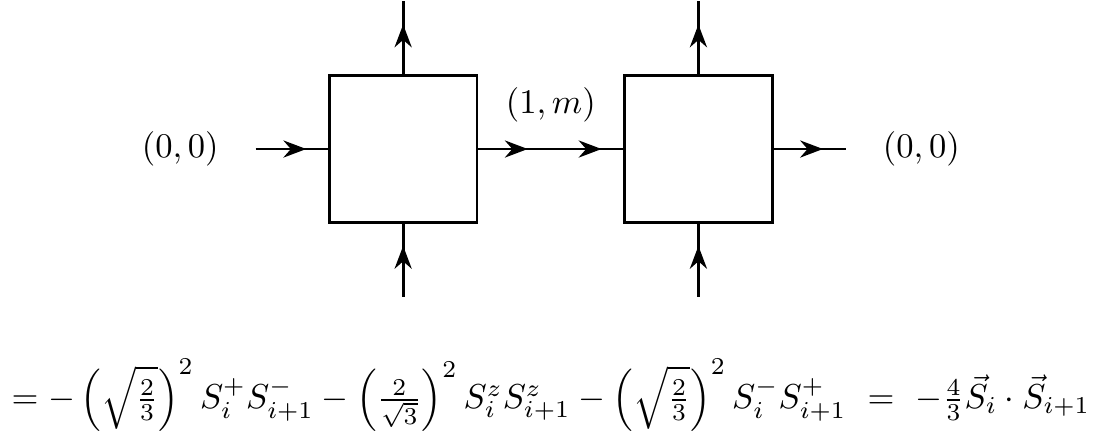}
	\caption{The contraction of the MPO tensors in Figs.~\ref{fig:TikZ_Files_SU(2)_MPO_SdotS_1} and \ref{fig:TikZ_Files_SU(2)_MPO_SdotS_2} produces the desired 2-site Heisenberg interaction with a $-4/3$ prefactor. The sum is over the values of $m$ for the spin-1 channel of the bond index.}
	\label{fig:TikZ_Files_MPO_Contraction_Heisenberg}
\end{figure}

Since the term $\mathbf{S} \cdot \mathbf{S}$ produces a scalar, we are interested in MPOs that transform as a scalar as well, which means that the bond indices at the left and right ends of the MPO need to have spin 0. For two physical spins $1/2$, it is easy to check that this implies that the connecting bond index between two MPO sites can only have either spin 0 or 1. While the spin 0 channel is trivial and corresponds to the application of the identity operator on the physical spins, the spin 1 channel \emph{necessarily} generates the dot product. This is because the only two scalar operators for two spins are $\mathbb{I} \otimes \mathbb{I}$ and $\mathbf{S} \cdot \mathbf{S} \equiv S^x \otimes S^x +  S^y \otimes S^y +  S^z \otimes S^z$. For the left and right MPO tensors, we can now write down all the coefficients once all the irreducible representations are fixed. Focusing on the spin-1 channel for the connecting index the results are shown in Fig.~\ref{fig:TikZ_Files_SU(2)_MPO_SdotS_1} and Fig.~\ref{fig:TikZ_Files_SU(2)_MPO_SdotS_2}. It is then straightforward to check that the contraction of the left and right MPO tensors from Figs.~\ref{fig:TikZ_Files_SU(2)_MPO_SdotS_1} and \ref{fig:TikZ_Files_SU(2)_MPO_SdotS_2} gives back the desired dot product $\mathbf{S} \cdot \mathbf{S}$ of the 2-site Heisenberg Hamiltonian, with a prefactor of $-4/3$ as shown in Fig.~\ref{fig:TikZ_Files_MPO_Contraction_Heisenberg}.

\begin{figure}
	\centering
	\includegraphics[width=\columnwidth]{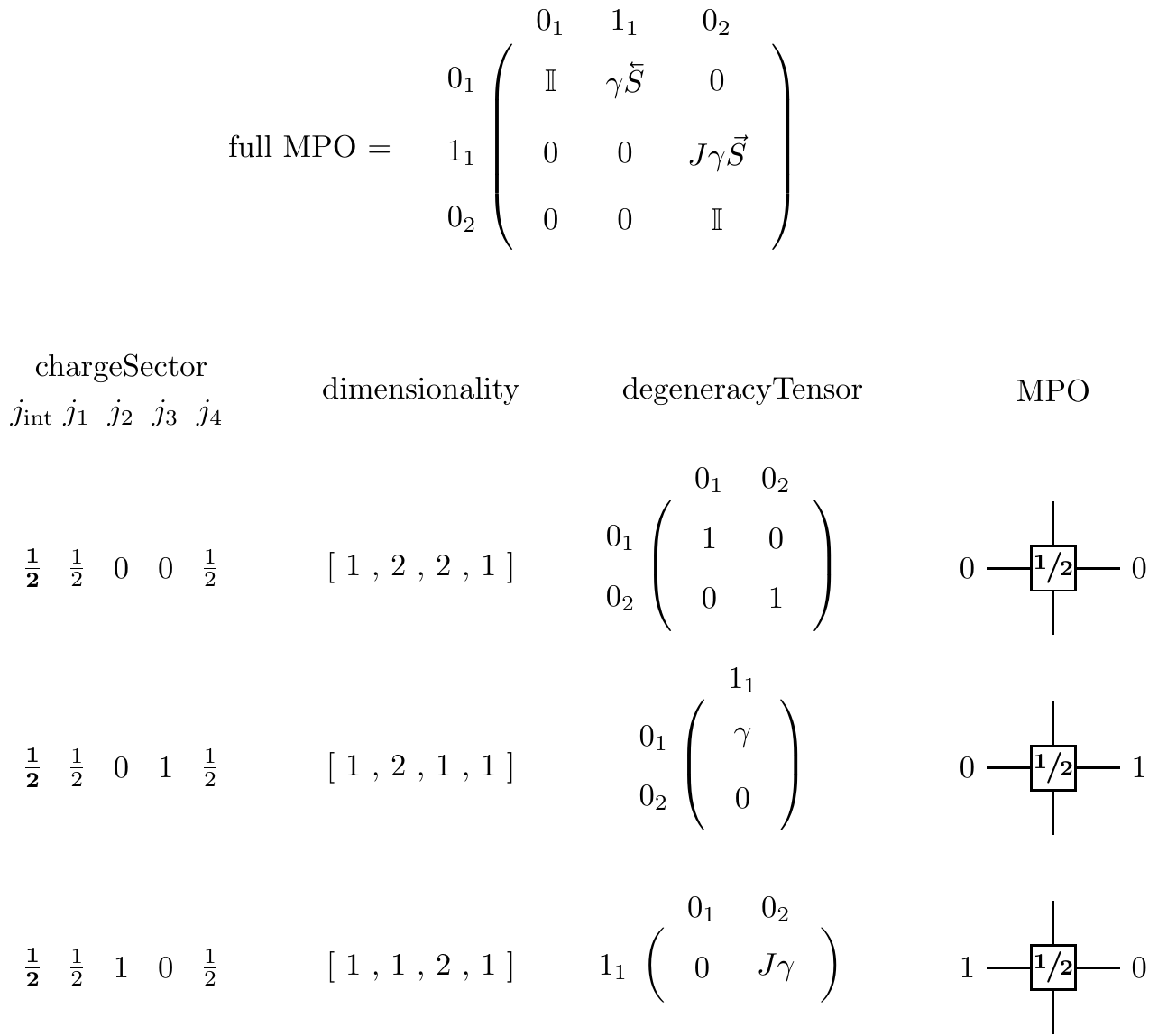}
	\caption{MPO tensor with SU(2) symmetry for the Heisenberg quantum spin chain. The degeneracy tensors go together with the rank-4 structural tensor for the spin sectors shown. All other spin sectors have vanishing degeneracy tensors. The internal spin for every block is shown in bold font, the physical spin is always $S = 1/2$. The labels of the MPO indices are in the order $D$, $L$, $R$, $U$.}
	\label{fig:TikZ_Files_Heisenberg_Model_SymmetricMPO_HeisenbergModel}
\end{figure}

Next, we can easily build the $SU(2)$-invariant MPO for the Heisenberg quantum spin chain with ${H = \sum_i \mathbf{S}_i \cdot \mathbf{S}_{i+1}}$ by adding one more spin zero channel to the MPO bond index. More specifically, the irrep with spin zero will have degeneracy two, i.e., we will have $0_1$ and $0_2$ (the subscript refers to the degeneracy). Combined with the non-degenerate spin channel $1_1$, the MPO can be written according to Fig.~\ref{fig:TikZ_Files_Heisenberg_Model_SymmetricMPO_HeisenbergModel}, where we show the details of the degeneracy tensors  accompanying the structural (Clebsch-Gordan) part. As in the non-symmetric part, the irrep $0_1$ ``propagates" through the bond indices until an interaction is hit (which is mediated by irrep $1$), and onwards propagates the irrep $0_2$. In describing this MPO tensor we have used the notation
\beq
\cev{S} = \begin{pmatrix} -\sqrt{\frac{2}{3}} S^+, & \frac{2}{\sqrt{3}} S^z, & \sqrt{\frac{2}{3}} S^-, \end{pmatrix}, ~
\vec{S} = \begin{pmatrix} \sqrt{\frac{2}{3}} S^- \\ -\frac{2}{\sqrt{3}} S^z \\ -\sqrt{\frac{2}{3}} S^+ \end{pmatrix},
\eeq
for a system of spin-$1/2$. Moreover, we also defined the factor $\gamma \equiv \rm i \sqrt{3/4}$ that compensates the unavoidable factor -4/3, which appears due to the contraction of the Clebsch-Gordan tensors as shown in Fig.~\ref{fig:TikZ_Files_MPO_Contraction_Heisenberg}.

\subsection{Chiral 3-spin interaction} 
\label{ssub:scalar_triple_product}
As a second example of a scalar operator we consider the triple product $\mathbf{S} \cdot (\mathbf{S} \times \mathbf{S})$. In this case, the MPO spans over three sites. The interaction on the second site can only be mediated by having a spin one representation on both virtual legs of the central MPO tensor. If this tensor had a spin zero representation on either leg, then it would not be possible to generate a three-body interaction. Moreover, a spin two representation is excluded since the 3-site MPO can then no longer be terminated by a spin zero after the third site, which we demand in order for it to be a scalar. Additionally, the central MPO tensor will have different internal spins $1/2$ and/or $3/2$, since
\beq
	\frac 1 2 \otimes 1 = \frac 1 2 \oplus \frac 3 2  ,
\eeq
where spin $1$ will come from some bond index, and spin $1/2$ from some physical index.

\begin{figure}
	\centering
	\includegraphics[width=\columnwidth]{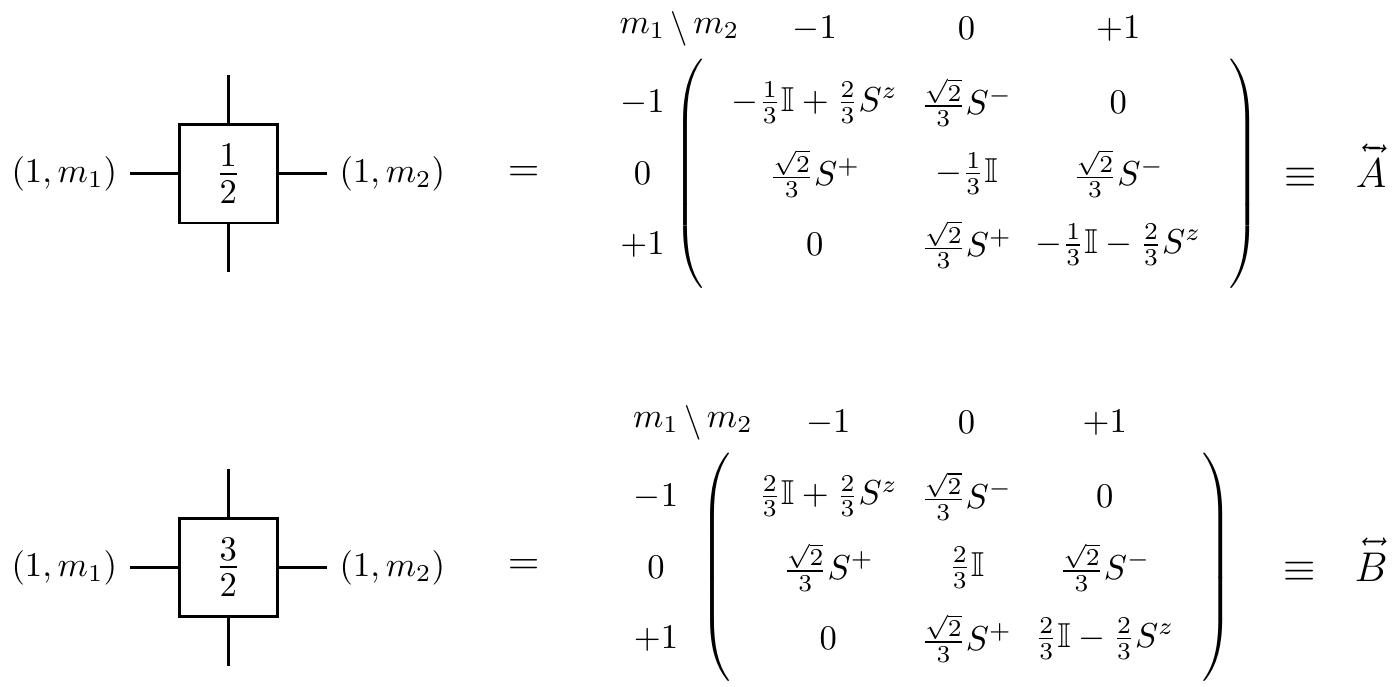}
	\caption{Possible coefficients of the central MPO tensor for the scalar triple product. Here the internal leg of the MPO can have the two spin representations $1/2$ and $3/2$.}
	\label{fig:TikZ_Files_Triangle_Model_Clebsch-Gordan_Matrices}
\end{figure}

In order to start and terminate the interaction in the MPO, we can reuse the terms of the Heisenberg interaction from Fig.~\ref{fig:TikZ_Files_SU(2)_MPO_SdotS_1} and Fig.~\ref{fig:TikZ_Files_SU(2)_MPO_SdotS_2}. The additional terms of the MPO for the triple product can be again constructed by evaluating the Clebsch-Gordan coefficients for fixed spin representations of the virtual legs for the central MPO tensor. These are given explicitly in Fig.~\ref{fig:TikZ_Files_Triangle_Model_Clebsch-Gordan_Matrices}, for the two possible values $1/2$ and $3/2$ of the internal leg.

\begin{figure}
	\centering
	\includegraphics[width=\columnwidth]{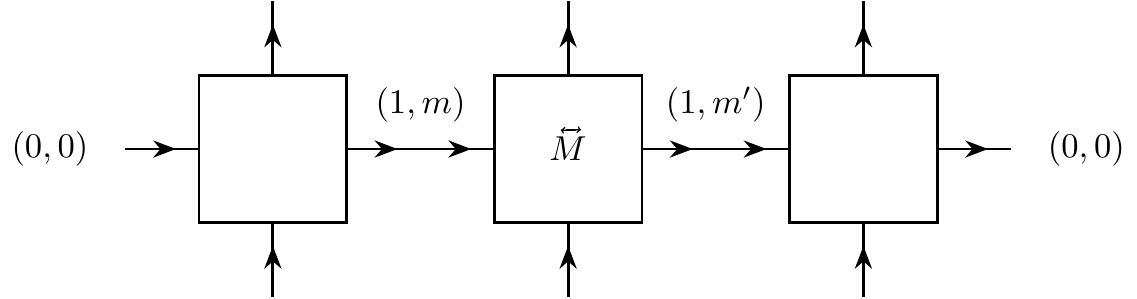}
	\caption{General construction of a three-site MPO with the linear combination $\protect\dvec{M} = \alpha \protect\dvec{A} + \beta \protect\dvec{B}$. Summation over the common indices $m$ and $m^\prime$ is assumed.}
	\label{fig:TikZ_Files_Triangle_Model_MPO_Contraction_Triangle}
\end{figure}

In order to construct the MPO we can now take linear combinations of matrices $A$ and $B$ in Fig.~\ref{fig:TikZ_Files_Triangle_Model_Clebsch-Gordan_Matrices} together with the left and right tensors from Fig.~\ref{fig:TikZ_Files_SU(2)_MPO_SdotS_1} and Fig.~\ref{fig:TikZ_Files_SU(2)_MPO_SdotS_2}. This is shown in \ref{fig:TikZ_Files_Triangle_Model_MPO_Contraction_Triangle}. We also use the notation
\beq
	\dvec{M} = \alpha \dvec{A} + \beta \dvec{B}  .
\eeq
Here the arrowheads indicate the center site of the MPO, in analogy to Fig.~\ref{fig:TikZ_Files_Heisenberg_Model_SymmetricMPO_HeisenbergModel}, where arrows were used to signal the start and end of the interaction.

The free parameters $\alpha$ and $\beta$ can now be chosen in order to reproduce the desired interaction. {As a first option one could choose the superposition $-A+B$. For this choice the MPO tensor at the central site simplifies to
\beq
	- A + B = \begin{pmatrix}
		\mathbb I 	& 0				& 0 		\\
		0			& \mathbb I		& 0			\\
		0			& 0				& \mathbb I \\
	\end{pmatrix} ,
\eeq
such that the overall three-body MPO reproduces a next-to-nearest-neighbour Heisenberg interaction ${\bf S_1} \cdot {\bf S_3}$ instead of the scalar triple product we aim for. As a matter of fact, this is also a valid scalar for three spins, where the second spin simply does not interact. By playing with this choice of $\alpha$ and $\beta$ it is also possible to construct MPOs for long-range interactions. In our case, though, we find that in order to generate the chiral triple product it is necessary to choose $\alpha = -i$ and $\beta = -i/2$, in which case the MPO tensor at the central site becomes
\beq
	-i\left( A + \frac 1 2 B \right)	= i\begin{pmatrix}
			-S^z 						& -\frac{\sqrt{2}}{2} S^-		& 0							\\
			-\frac{\sqrt{2}}{2} S^+		& 0								& -\frac{\sqrt{2}}{2} S^-	\\
			0							& -\frac{\sqrt{2}}{2} S^+		& S^z
		\end{pmatrix} .
\eeq
Evaluating the sum over the three MPO tensors in Fig.~\ref{fig:TikZ_Files_Triangle_Model_MPO_Contraction_Triangle} yields
\beqa
	H &=& -\frac{4i}{6} ( S^z S^+ S^- - S^z S^- S^+ + S^+ S^- S^z \nonumber \\
	&-& S^+ S^z S^- + S^- S^z S^+ - S^- S^+ S^z)	\nonumber \\
	  &=& -\frac 4 3 {\bf S}_1 \cdot \left( {\bf S}_2 \times {\bf S}_3 \right) .
\eeqa
Notice that here again the factor $-4/3$ appears due to the relation between spin-$1/2$ operators and Clebsch-Gordan coefficients. From these tensors it is then easy to construct an MPO for a Hamiltonian made of a sum of chiral 3-spin interactions.

\subsection{2-spin and 3-spin interactions together}

\begin{figure}
	\centering
	\includegraphics[width=\columnwidth]{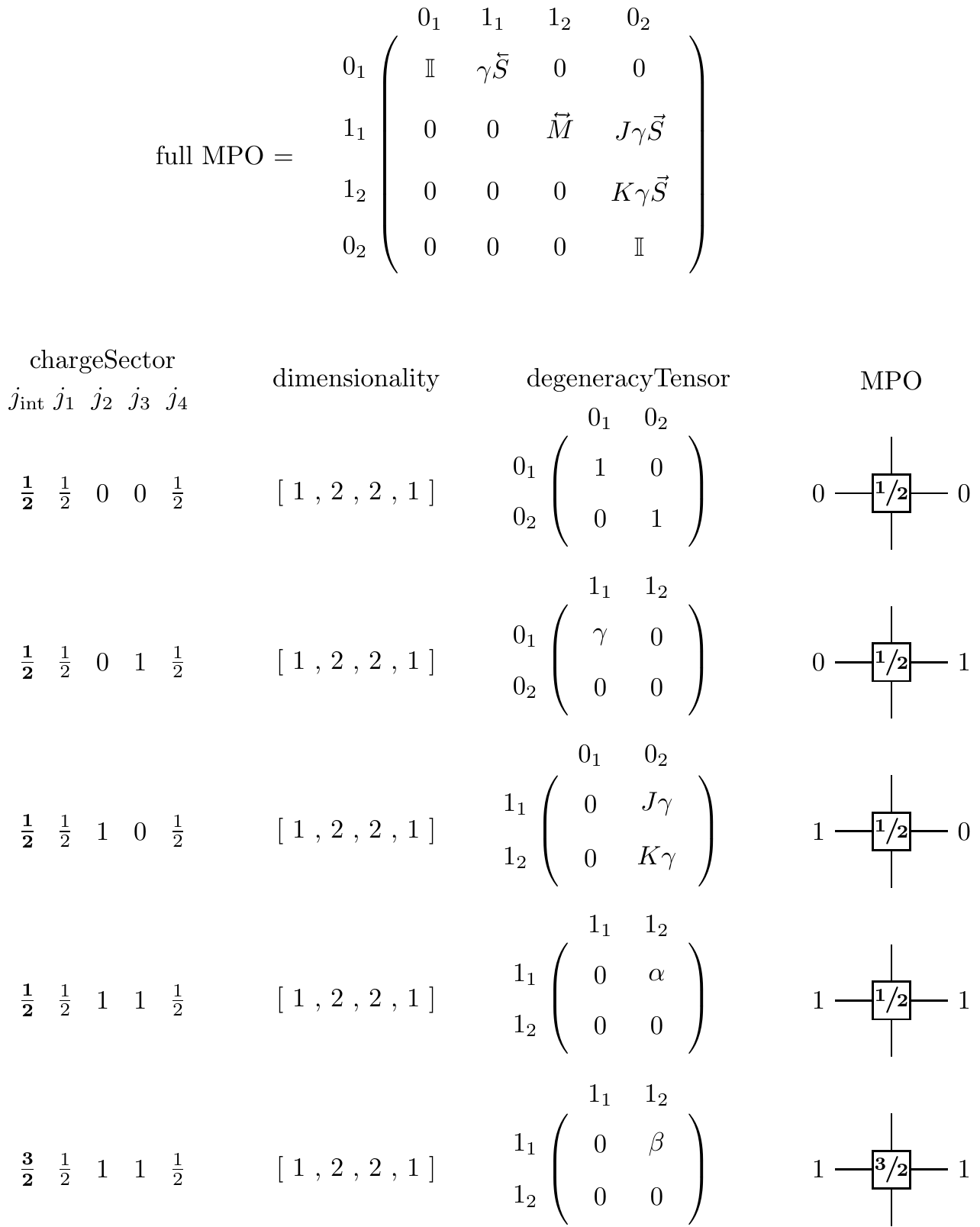}
	\caption{MPO tensor with $SU(2)$ symmetry for the quantum spin chain Heisenberg and chiral 3-spin interactions. The degeneracy tensors go together with the rank-4 structural tensor for the spin sectors shown. All other spin sectors have vanishing degeneracy tensors. The operator $\protect\dvec{M}$ is the combination of the operators $\protect\dvec{A}$ and $\protect\dvec{B}$ with proper weights. Again the labels of the MPO indices are in the order D,L,R,U.}
	\label{fig:TikZ_Files_Triangle_Model_SymmetricMPO_HeisenbergModel}
\end{figure}

\begin{figure}
	\centering
	\includegraphics[width=0.325\columnwidth]{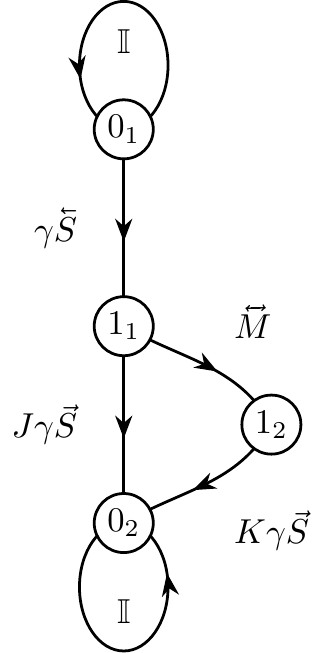}
	\caption{Finite State Machine for the MPO of the spin-1/2 nearest-neighbour Heisenberg model with chiral 3-spin interactions. The different states correspond to the different spin sectors for the bond dimensions of the MPO.}
	\label{fig:TikZ_Files_SU(2)_Hamiltonian_FSM_TriangleModel}
\end{figure}
\begin{figure}
	\centering
	\includegraphics[width=0.8\columnwidth]{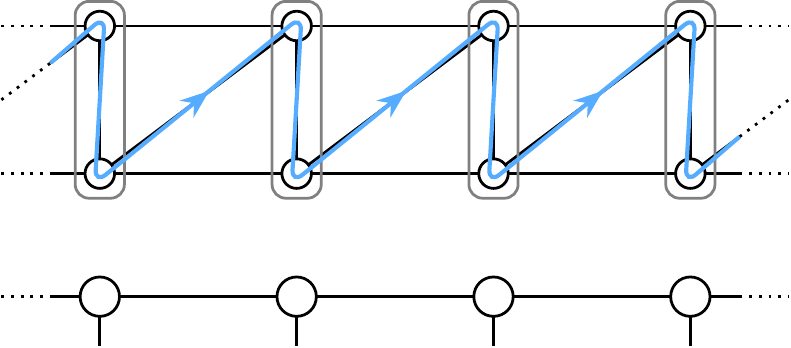}
	\caption{Snake pattern for the MPO and also coarse-graining of two ladder sites into one site of the MPS.}
	\label{snake}
\end{figure}

We can now construct an MPO for a Hamiltonian such as
\beq
H = J \sum_i {\bf S}_{i} {\bf S}_{i+1} + K \sum_i {\bf S}_i \cdot \left( {\bf S}_{i+1} \times {\bf S}_{i+2} \right)  ,
\eeq
i.e., a sum of 2-spin Heisenberg interactions and chiral 3-spin interactions. This Hamiltonian is an $SU(2)$ scalar, because it is constructed as a sum of scalar operators. $J$ and $K$ are parameters giving more weight to one term or the other.

As in the case of the plain Heisenberg model, we need two spin zero sectors in the bond dimensions of the MPO that take care of applying the identity to all sites to the left and to the right of the interacting sites. Moreover, here we also need two spin one sectors in the MPO bond dimension: one mediating the 2-spin interaction, and the other mediating the 3-spin interaction. The resulting MPO tensor is given in Fig.~\ref{fig:TikZ_Files_Triangle_Model_SymmetricMPO_HeisenbergModel}, where we specify the structural part, corresponding to the Clebsch-Gordan coefficients,  as well as the degeneracy part. The structure of the MPO can also be represented by a \emph{Finite State Machine}, as shown in Fig.~\ref{fig:TikZ_Files_SU(2)_Hamiltonian_FSM_TriangleModel}.

\subsection{Chiral 3-spin interactions on the ladder}

For the purposes of this paper, we simulated a Hamiltonian with chiral 3-spin interactions of the triangles of the ladder in Fig.~\ref{fig1}, in which we considered alternating orientations for the triangles as explained above. In order to construct an MPO, we considered the snake pattern from Fig.~\ref{snake}, and applied the techniques discussed previously to construct an MPO for the sum of the different 3-spin interactions. Then, as shown in the figure, we coarse-grained the two spins for each rung of the ladder into a single physical site (with irreps $0_1 \oplus 1_1$). In this way, the resulting MPO has a 2-site unit cell and is the one used in the iDMRG algorithm with a 2-site update.

\begin{figure}
	\centering
	\includegraphics[width=\columnwidth]{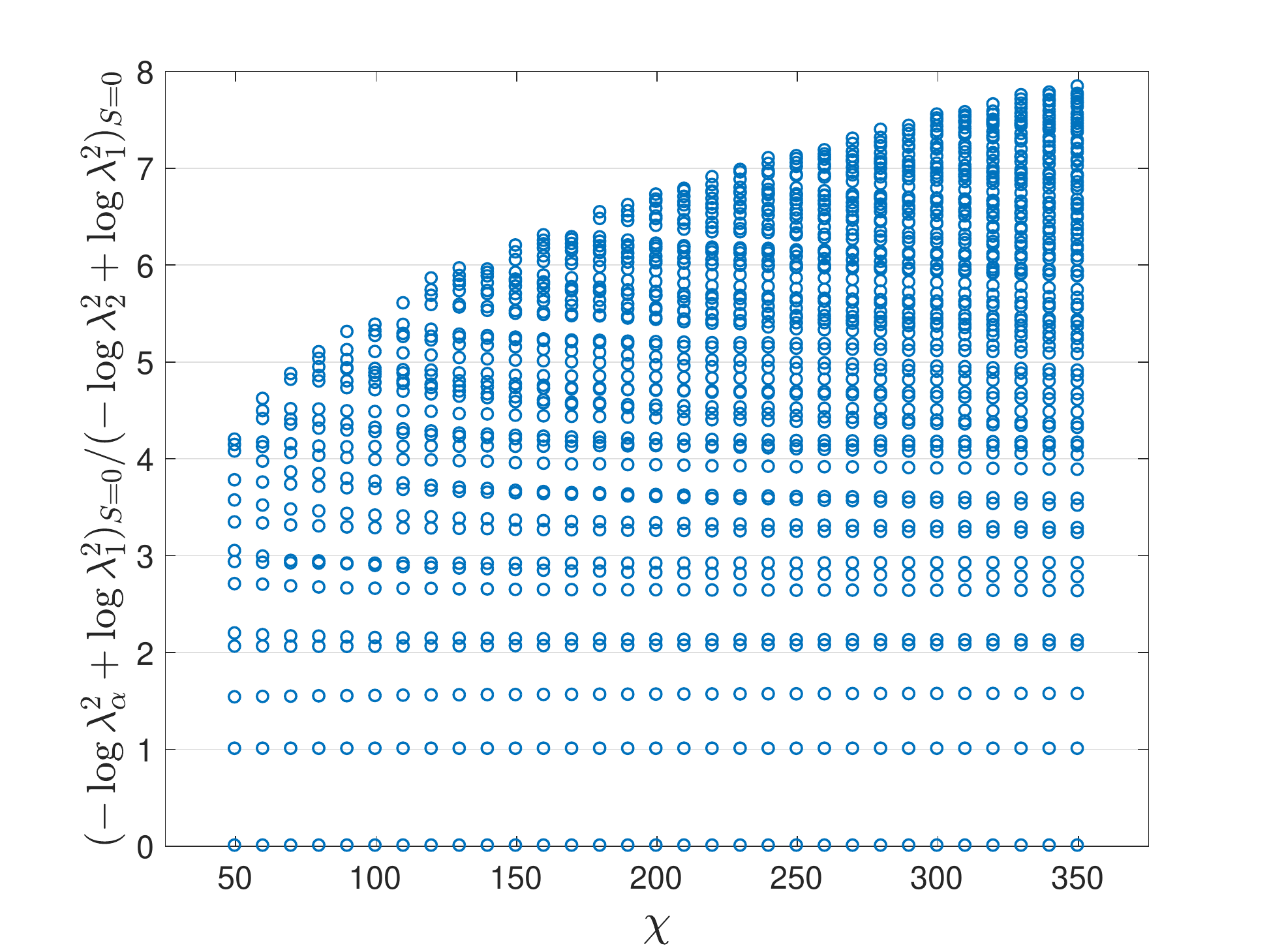}
	\caption{Scaling of the entanglement spectrum for $S=0$ with the MPS symmetric bond dimension $\chi$, in normalized units.}
	\label{EScal0}
\end{figure}

\section{Finite-entanglement scaling of the entanglement spectrum}
\label{append4}

In this appendix we show our results for the scaling with the bond dimension $\chi$ of the entanglement spectrum for the different spin sectors $S$. Our results are shown in Figs.~\ref{EScal0} - \ref{EScal3} for integer spins $S=0$ up to $S=3$. Each dot in the plots is a $(2S+1)$-plet. As we can see, the lowest part of the entanglement spectrum converges quickly with the bond dimension. The distribution of the lowest-lying entanglement energies tends to have an equidistant structure, typical of a CFT. The values for the largest possible bond dimension, which we take as essentially converged for the lowest-lying part of the spectrum, correspond to the ones shown in Fig.~\ref{entanglementSpectrum_1}. We also notice that the convergence of the individual entanglement energies with the symmetric bond dimension seems to be algebraic as opposed to exponential. In practice, this means that from our plots one can extract the behavior
\beq
\varepsilon_\alpha \approx \frac{1}{\chi^{\mu}},
\eeq
for the $\alpha$th entanglement energy $\varepsilon_\alpha$, with $\mu$ an exponent controlling the behavior at large $\chi$. According to our data, the exponent $\mu$ may depend on the index $\alpha$ itself, i.e., be different for each one of the entanglement energies. Even if purely empirical, this behavior seems to hold well for all the studied values of the spin $S$. According to the results presented in this paper, we take this also as a strong  indication that the system is critical and has an infinite correlation length.

\begin{figure}
	\centering
	\includegraphics[width=\columnwidth]{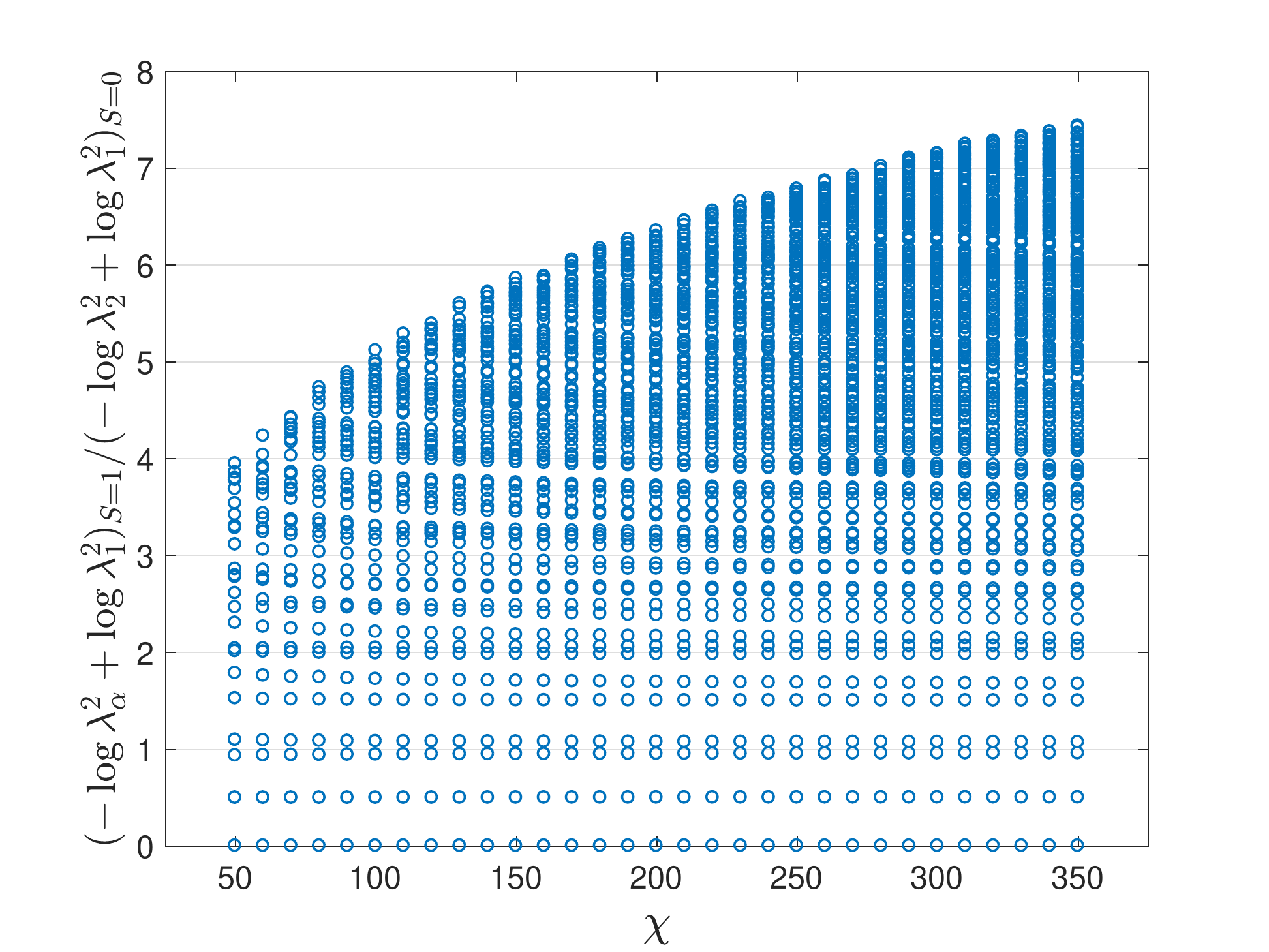}
	\caption{Scaling of the entanglement spectrum for $S=1$ with the MPS symmetric bond dimension $\chi$, in normalized units.}
	\label{EScal1}
\end{figure}
\begin{figure}
	\centering
	\includegraphics[width=\columnwidth]{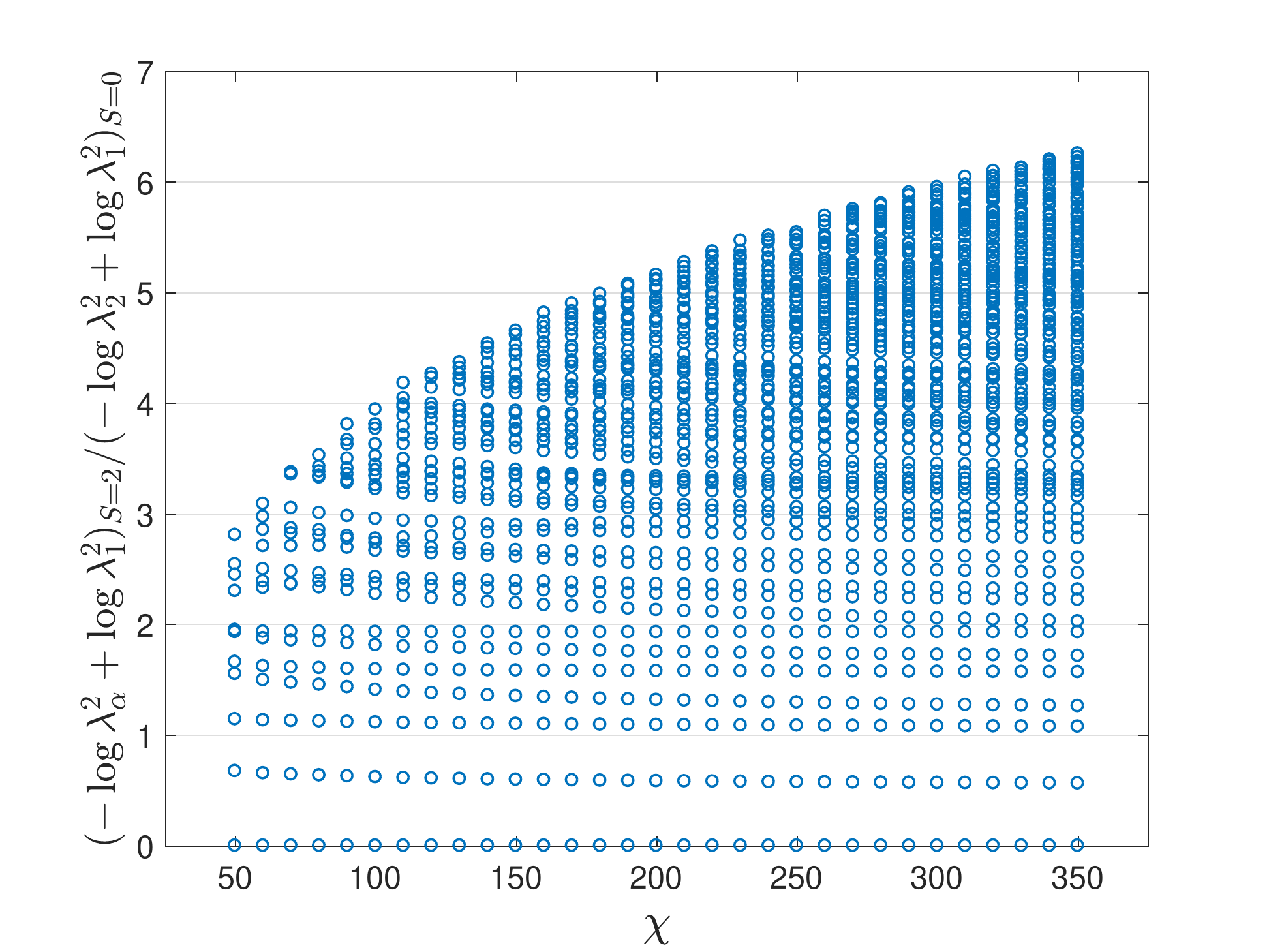}
	\caption{Scaling of the entanglement spectrum for $S=2$ with the MPS symmetric bond dimension $\chi$, in normalized units.}
	\label{EScal2}
\end{figure}
\begin{figure}
	\centering
	\includegraphics[width=\columnwidth]{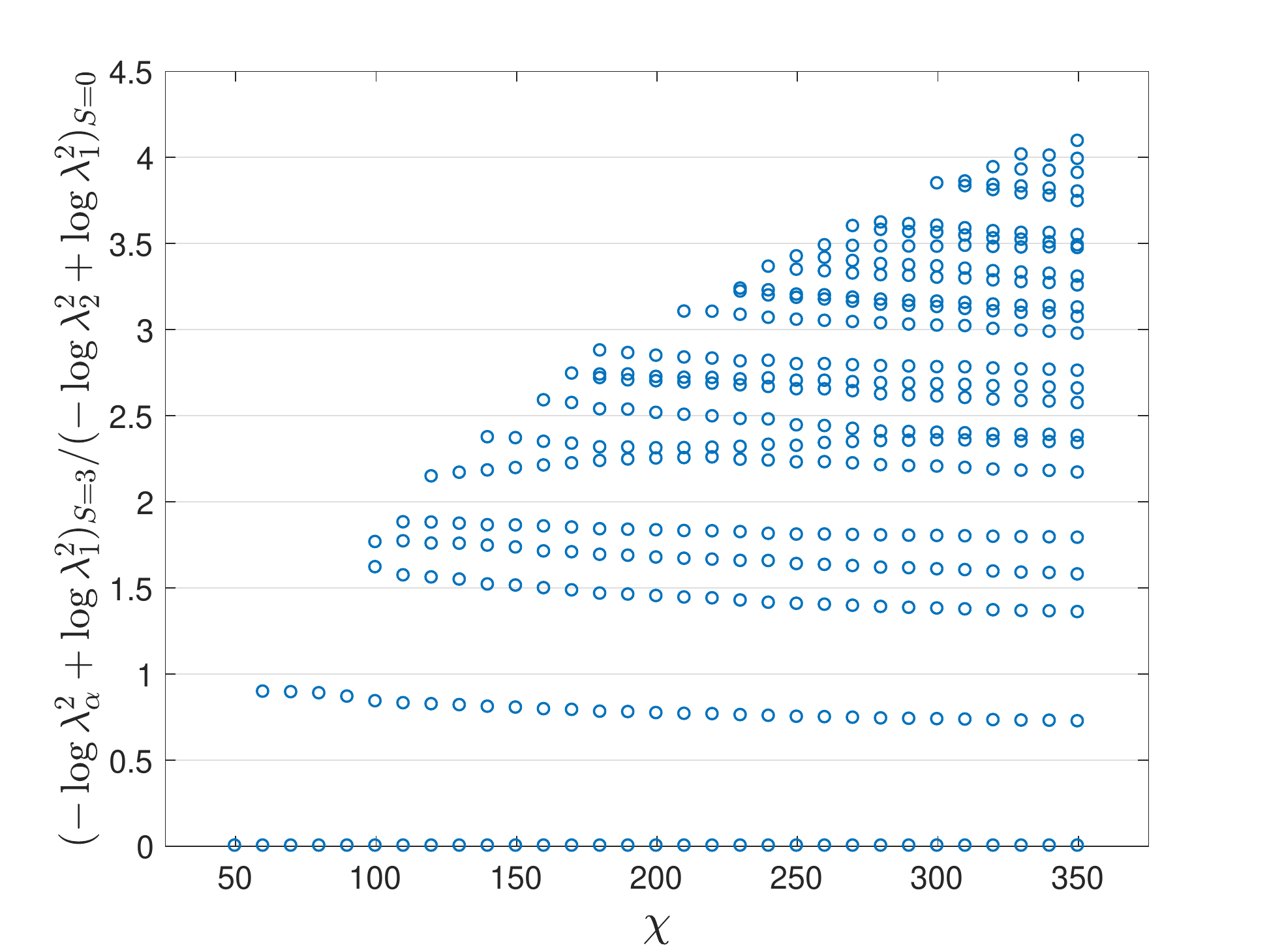}
	\caption{Scaling of the entanglement spectrum for $S=3$ with the MPS symmetric bond dimension $\chi$, in normalized units.}
	\label{EScal3}
\end{figure}

 \end{document}